\documentclass[useAMS,usenatbib]{mn2e}
\usepackage{epsfig} % to have figures in the .eps format
\usepackage{epstopdf}
\usepackage{graphicx}
\usepackage{amsmath} %math stuff
\usepackage{hyperref}
\usepackage{times}
\usepackage[T1]{fontenc}
\usepackage{tabularx}
\usepackage{abbreviations}
\usepackage{subfig}

\usepackage{breakcites}

\newcommand{\HII}{\mbox{H II}}
\newcommand{\OIII}{\mbox{[O III]}}
\newcommand{\OII}{\mbox{[O II]}}

\newcommand{\NII}{\mbox{[N II]}}
\newcommand{\SII}{\mbox{[S II]}}
\newcommand{\Ha}{H$\alpha$}
\newcommand{\Hb}{H$\beta$}

\newcommand{\NIIHa}{\NII/H$\alpha$}

\newcommand{\OIIIHb}{\OIII/H$\beta$}

\begin{document}

\title[Separating Star Formation and AGN Activity]{Dissecting Galaxies: Spatial and Spectral Separation of Emission Excited by Star Formation and AGN Activity}

\author[R. L. Davies et al.]{Rebecca L. Davies$^{1}$\thanks{E-mail:
Rebecca.Davies@anu.edu.au}, Brent Groves$^{1}$, Lisa J. Kewley$^{1}$, Michael A. Dopita$^{1}$, \newauthor Elise J. Hampton$^1$, Prajval Shastri$^2$, Julia Scharw\"achter$^3$, Ralph Sutherland$^1$, \newauthor Preeti Kharb$^2$, Harish Bhatt$^2$, Chichuan Jin$^4$, Julie Banfield$^{1,5}$, Ingyin Zaw$^6$,\newauthor Bethan James$^7$, St\'ephanie Juneau$^8$,  \& Shweta Srivastava$^{9}$ \\ \\
$^{1}$Research School of Astronomy and Astrophysics, Australian National University, Canberra, ACT 2611, Australia \\
$^2$Indian Institute of Astrophysics, Sarjapur Road, Bengaluru 560034, India \\
$^3$LERMA, Observatoire de Paris, PSL, CNRS, Sorbonne Universit\'es, UPMC, F-75014 Paris, France \\
$^4$Max-Planck-Institut f\"{u}r Extraterrestrische Physik, Giessenbachstrasse, D-85748 Garching, Germany \\
$^5$ARC Centre of Excellence for All-Sky Astrophysics (CAASTRO) \\
$^6$New York University (Abu Dhabi) , 70 Washington Sq. S, New York, NY 10012, USA \\
$^7$Institute of Astronomy, Cambridge University, Madingley Road, Cambridge CB3 0HA, UK \\
$^8$CEA-Saclay, DSM/IRFU/SAp, 91191 Gif-sur-Yvette, France \\
$^{9}$Astronomy and Astrophysics Division, Physical Research Laboratory, Ahmedabad 380009, India} 

\maketitle

\begin{abstract}
The optical spectra of Seyfert galaxies are often dominated by emission lines excited by both star formation and AGN activity. Standard calibrations (such as for the star formation rate) are not applicable to such composite (mixed) spectra. In this paper, we describe how integral field data can be used to spectrally and spatially separate emission associated with star formation from emission associated with accretion onto an active galactic nucleus (AGN). We demonstrate our method using integral field data for two AGN host galaxies (NGC~5728 and NGC~7679) from the Siding Spring Southern Seyfert Spectroscopic Snapshot Survey (S7). The spectra of NGC~5728 and NGC~7679 form clear sequences of AGN fraction on standard emission line ratio diagnostic diagrams. We show that the emission line luminosities of the majority ($>$~85 per cent) of spectra along each AGN fraction sequence can be reproduced by linear superpositions of the emission line luminosities of one AGN dominated spectrum and one star formation dominated spectrum. We separate the \Ha, \Hb, \NII$\lambda$6583, \mbox{\SII$\lambda \lambda$6716, 6731}, \OIII$\lambda$5007 and \mbox{\OII$\lambda \lambda$3726, 3729} luminosities of every spaxel into contributions from star formation and AGN activity. The decomposed emission line images are used to derive the star formation rates and AGN bolometric luminosities for NGC~5728 and NGC~7679. Our calculated values are mostly consistent with independent estimates from data at other wavelengths. The recovered star forming and AGN components also have distinct spatial distributions which trace structures seen in high resolution imaging of the galaxies, providing independent confirmation that our decomposition has been successful.
\end{abstract}


\section{Introduction}
The impact of AGN activity on the evolution of the host galaxy is a topic of significant debate. In the local universe, AGN reside in galaxies at a range of evolutionary stages, from spiral galaxies with strong circumnuclear star formation \citep[e.g.][]{GonzalezDelgado01, Joguet01, StorchiBergmann01, Raimann03, CidFernandes04, Gu06, Davies07}, to `green valley' galaxies (whose optical colours suggest that they may be transitioning from the star-forming blue cloud to the quiescent red sequence) \citep[e.g.][]{Ka03,Baldry04,Schawinski10, Leslie15} and quiescent elliptical galaxies \citep[e.g.][]{Olsen13}. The evolved nature of many AGN host galaxies may be a direct result of AGN feedback quenching star formation \citep[e.g.][]{DiMatteo05,Nandra07, Schawinski09, Schawinski10, Leslie15} or may simply be a consequence of the increase in both the fraction of passive galaxies and the fraction of galaxies hosting AGN with increasing stellar mass \citep[e.g.][]{Ka03,Baldry04}. Global colours and integrated star formation rates (SFRs) provide limited insight into the physical processes impacting the gas reservoirs within AGN host galaxies, making it difficult to distinguish between these scenarios.

The combination of imaging and spectroscopy can provide a more direct view of the impact of AGN activity on the surrounding interstellar medium (ISM). For example, \citet{Cresci15} used near infrared integral field spectroscopy (IFS) to identify both positive and negative AGN feedback in a radio quiet quasar at \mbox{z = 1.59}. They find that star formation is being suppressed due to the entrainment of molecular gas within an AGN driven outflow, but is also being triggered at the edges of the outflow due to the pressure imposed on the surrounding ISM. Several other spatially resolved and/or multi-wavelength studies have found direct evidence for star formation being suppressed due to strong outflows \citep[e.g.][]{CanoDiaz12} or triggered due to the compression of molecular clouds by radio jets \citep[e.g.][]{Croft06, Elbaz09, Rauch13, Salome15}. These studies highlight the power of spatial information for providing insights into the connection between star formation and AGN activity in galaxies. 

Modern IFS surveys (such as S7 \citep{Dopita15}, the Sydney AAO Multi-Object Integral Field Spectrograph (SAMI) Survey \citep{Croom12, Bryant15}, the Calar Alto Legacy Integral Field Area (CALIFA) Survey \citep{Sanchez12} and the Mapping Nearby Galaxies at Apache Point Observatory (MaNGA) Survey \citep{Bundy15}) are providing optical spectra for many tens of spatially resolved regions across hundreds to thousands of galaxies. These data can potentially be used to map the SFR and the strength of the AGN ionizing radiation field across large samples of AGN host galaxies. However, extracting this information from the spectra of Seyfert galaxies (which often have significant contributions from both star formation and AGN activity) is non-trivial. The \Ha\ luminosity of an \HII\ region is directly proportional to the SFR \citep[e.g.][]{Kennicutt98}. However, \Ha\ can be collisionally excited in the presence of an AGN ionizing radiation field, and therefore the \Ha\ luminosity is not a valid diagnostic of the SFR in AGN narrow line regions (NLRs). Similarly, the \OIII\ $\lambda$5007 luminosity is directly proportional to the strength of the local AGN ionizing radiation field within AGN dominated regions \citep[e.g.][]{Heckman04, Lamastra09}, but is not a valid diagnostic of the radiation field strength in regions with active star formation. It is therefore necessary to separate the composite spectra of Seyfert galaxies into contributions from star formation and AGN activity before calculating SFRs or AGN luminosities.

Many techniques exist for separating emission associated with star formation and AGN activity. In the mid infrared, these two ionization mechanisms have very different spectral signatures and can be separated using template fitting \citep[e.g.][]{Nardini08, AlonsoHerrero12, Kirkpatrick14, HernanCaballero15}. However, mid infrared observations typically have low spatial resolution and therefore cannot provide detailed maps of the emission associated with star formation and AGN activity across galaxies. On the other hand, optical IFS surveys represent a very promising avenue for studying the connection between star formation and AGN activity on scales of hundreds of parcsecs to kiloparsecs. The optical spectra of AGN host galaxies are dominated by strong emission lines which can be excited by both emission from massive stars and the AGN ionizing radiation field. The ratios of forbidden to recombination lines are dependent on the relative contributions of star formation and AGN activity and can therefore be used to estimate the fraction of the line emission excited by AGN activity (`AGN fraction'). 

The \NIIHa\ vs. \OIIIHb\ diagnostic diagram is a valuable tool for separating emission associated with star formation and AGN activity \citep{Baldwin81, Veilleux87, Ke01a}. Spectra dominated by star formation fall along the star-forming sequence which traces variations in the ionized gas abundance \citep{Dopita86, Dopita00}. Spectra with contributions from more energetic ionization mechanisms lie along the AGN branch of the diagnostic diagram which spans from the enriched end of the star-forming sequence towards larger \NIIHa\ and \OIIIHb\ ratios. The presence of a harder ionizing radiation field increases the collisional excitation rate in the nebula and therefore increases the strengths of forbidden lines such as \NII\ and \OIII\ (produced by radiative transitions from collisionally excited metastable states) relative to the \Ha\ and \Hb\ recombination lines. The greater the AGN fraction, the greater the enhancement in the \NIIHa\ and \OIIIHb\ ratios and the further along the AGN branch a galaxy will lie. 

The Sloan Digital Sky Survey (SDSS) provided single fibre optical spectra for hundreds of thousands of galaxies in the local universe, leading to several pioneering works in the separation of star formation and AGN activity. \citet{Heckman04} corrected the \OIII\ luminosities of observed composite spectra for the contribution of star formation using average AGN fractions calculated for synthetic composite spectra (generated by summing observed star formation and AGN dominated spectra) in bins of \OIII\ luminosity. \citet{Ke06} established the `star-forming distance' ($d_{SF}$, the distance of a galaxy spectrum from the star-forming sequence of the \NIIHa\ vs. \OIIIHb\ diagnostic diagram) as a metric for the relative contribution of star formation to the line emission. Following this, \citet{Kauffmann09} used the positions of galaxies along the AGN branch of the diagnostic diagram to estimate AGN fractions and correct the \OIII\ luminosities for the contribution of star formation. These pioneering techniques facilitated the first large statistical studies of black hole accretion rates and Eddington ratios as a function of host galaxy properties.

With the advent of integral field spectropscopy, it is now possible to use similar techniques to separate star formation and AGN activity in individual spatially resolved regions within Seyfert host galaxies. \citet{Davies14b,Davies14a} showed that spectra extracted from individual spectral pixels (spaxels) of AGN host galaxies sometimes fall along tight sequences spanning the full range of line ratios observed along the SDSS global mixing sequence. The line ratios vary smoothly with galactocentric distance, from AGN-like line ratios in the galaxy nuclei to \HII-like line ratios at larger radii, providing strong evidence that the line ratio variations are primarily driven by variations in the AGN fraction \citep[see also][]{Scharwaechter11, Dopita14, Belfiore15, Davies16RP}. We note that the diagnostic line ratios are sensitive to the physical conditions of the ionized gas \citep[see e.g.][]{Ke03, Groves04, Dopita13} and therefore radial variations in the local ionization parameter and/or gas pressure would also drive radial variations in the line ratios \citep[e.g.][]{Greene11,Liu13}. However, the presence of spectra in the star forming, composite and AGN dominated regions of the \NIIHa\ vs. \OIIIHb\ diagnostic diagram within a single galaxy is inconsistent with ionization by a single mechanism and can only be explained by mixing. The observation of spectra at a range of AGN fractions makes it possible to directly constrain the shape of the starburst-AGN mixing curve for individual galaxies, without any \textit{a priori} knowledge of the ISM conditions. 

In this paper, we present a new method for separating emission associated with star formation and AGN activity in individual emission lines within individual spaxels of IFS datacubes. We describe our data and introduce our galaxy sample in Section \ref{sec:obs}. We discuss the motivation for our method and outline the steps in Section \ref{sec:method}. In Section \ref{sec:results} we present the results of the decomposition for each galaxy, and compare our results to independent tracers of star formation and AGN activity at other wavelengths. We discuss the advantages and limitations of our decomposition method and compare it to existing methods in Section \ref{sec:discussion}. We summarise our main conclusions in Section \ref{sec:conc}.

Throughout this paper we adopt cosmological parameters $H_{0} = 70.5 \, {\rm kms}^{-1}{\rm Mpc}^{-1}$, ${\Omega}_{\Lambda} = 0.73$, and $\Omega_{M}=0.27$ based on the 5-year \emph{Wilkinson Microwave Anisotropy Probe} (WMAP) results by \citet{Hinshaw09} and consistent with flat $\Lambda$-dominated cold dark matter ($\Lambda$ CDM) cosmology. 

\section{Observations, Data Analysis and Sample}
\label{sec:obs}
\subsection{Observations and Data Reduction}
Our data are drawn from S7, an integral field survey of 130 low redshift (\mbox{z $<$ 0.02}), southern (\mbox{declination $<$ 10$^\circ$}) AGN host galaxies \citep{Dopita15}. We observe the central \mbox{38 $\times$ 25} arcsec$^2$ region of each galaxy using the Wide Field Spectrograph (WiFeS; \citealt{Dopita07, Dopita10}) on the ANU 2.3m telescope, with a spatial sampling of 1 arcsec pixel$^{-1}$. WiFeS is a double-beam spectrograph with three gratings mounted on each arm (one high resolution grating; R~=~7000, \mbox{$\Delta v$ = 43 km s$^{-1}$} and two lower resolution gratings; R~=~3000, \mbox{$\Delta v = 100$ km s$^{-1}$}). Our S7 observations are conducted using the high resolution red grating (R7000, $\lambda$~=~530-\mbox{710 nm}) and a lower resolution blue grating (B3000, $\lambda$~=~340-\mbox{570 nm}). This grating combination provides us with both the high spectral resolution required to probe the detailed kinematics of the S7 galaxies and the wavelength coverage needed to observe all of the diagnostic emission lines in the blue region of the spectrum (\OII$\lambda \lambda$3727,3729, \Hb$\lambda$4861 and \OIII$\lambda \lambda$4959,5007). Full details of the observations can be found in \citet{Dopita15}.

The data were reduced using the Python pipeline \textsc{PyWiFeS} which performs overscan and bias subtraction, interpolates over bad CCD columns, removes cosmic rays, derives a wavelength solution from the arc lamp observations, performs flat-fielding to account for pixel-to-pixel sensitivity variations across the CCDs, re-samples the data into a cube and finally performs telluric correction and flux calibration. Absolute photometric calibration of the data cubes was performed using the STIS spectrophotometric standard stars\footnote{Fluxes available at: \newline {\url{www.mso.anu.edu.au/~bessell/FTP/Bohlin2013/GO12813.html}}}. The wavelength solutions are accurate to $\sim$0.05\AA\ (RMS) across the entire detector \citep{Childress14}, and the typical absolute spectrophotometric accuracy of the flux calibration is 4 per cent. \textsc{PyWiFeS} outputs both a data cube and an error cube for each galaxy. The \textsc{PyWiFeS} errors are propagated through all subsequent analysis to produce the errors on the final results.

\subsection{Emission Line Fitting}
\label{subsec:line_fit}
We use the IDL emission line fitting toolkit \textsc{LZIFU} (Ho et al., submitted, see \citealt{Ho14} for a brief description of the code) to fit the spectrum of each spaxel as a linear superposition of stellar templates and Gaussian emission line components.  \textsc{LZIFU} feeds each spectrum into the penalized pixel fitting routine (\textsc{pPXF}; \citealt{Cappellari04}) which fits the stellar continuum emission as a linear combination of the high spectral resolution (\mbox{$\Delta \lambda$ = 0.3\AA}) stellar templates from \citet{GonzalezDelgado05}. The errors output by \textsc{PyWiFeS} are taken into account when calculating the penalized likelihood of each template combination. The stellar continuum fit is subtracted from the original spectrum, leaving an emission line spectrum. \textsc{LZIFU} extracts the emission line fluxes and kinematics from the emission line spectra using the Levenberg-Marquardt least squares fitting routine \textsc{MPFIT} \citep{Markwardt09}. Each emission line spectrum is fit three times independently - once with a single component, once with two components and once with three components. Each component consists of a set of Gaussian functions (one associated with each emission line) which all have the same width (velocity dispersion) and velocity offset (relative to the velocity corresponding to the input spectroscopic redshift). The presence of multiple kinematic components within a single spectrum arises when emission associated with gas clouds of differing velocities is superimposed along the line of sight. An artificial neural network (ANN) is used to determine how many components are required for each spectrum, as described in Section \ref{subsec:ANN}.

The typical continuum fitting residuals are less than 5 per cent, and such small errors will generally have a negligible impact on the derived emission line fluxes. The largest uncertainty introduced by the stellar continuum fitting is in the Balmer absorption correction, which can significantly impact the measured fluxes for low equivalent width \Ha\ and \Hb\ lines. The Seyfert host galaxies being analysed in this paper display strong, high equivalent width emission lines (see e.g. spectra in Figure \ref{fig:galaxy_properties}). The Balmer absorption correction should not have a significant impact on the derived Balmer line fluxes in the galaxies of interest, and therefore the continuum fitting errors are not included in the error budget. \textsc{MPFIT} propagates the \textsc{PyWiFeS} errors through the least squares minimisation algorithm to produce the final error on the flux in each kinematic component of each emission line in each spaxel.

\begin{figure*}
\captionsetup[subfigure]{labelformat=empty}
\subfloat[]{\includegraphics[scale = 0.22, clip = true, trim = 5 5 5 5]{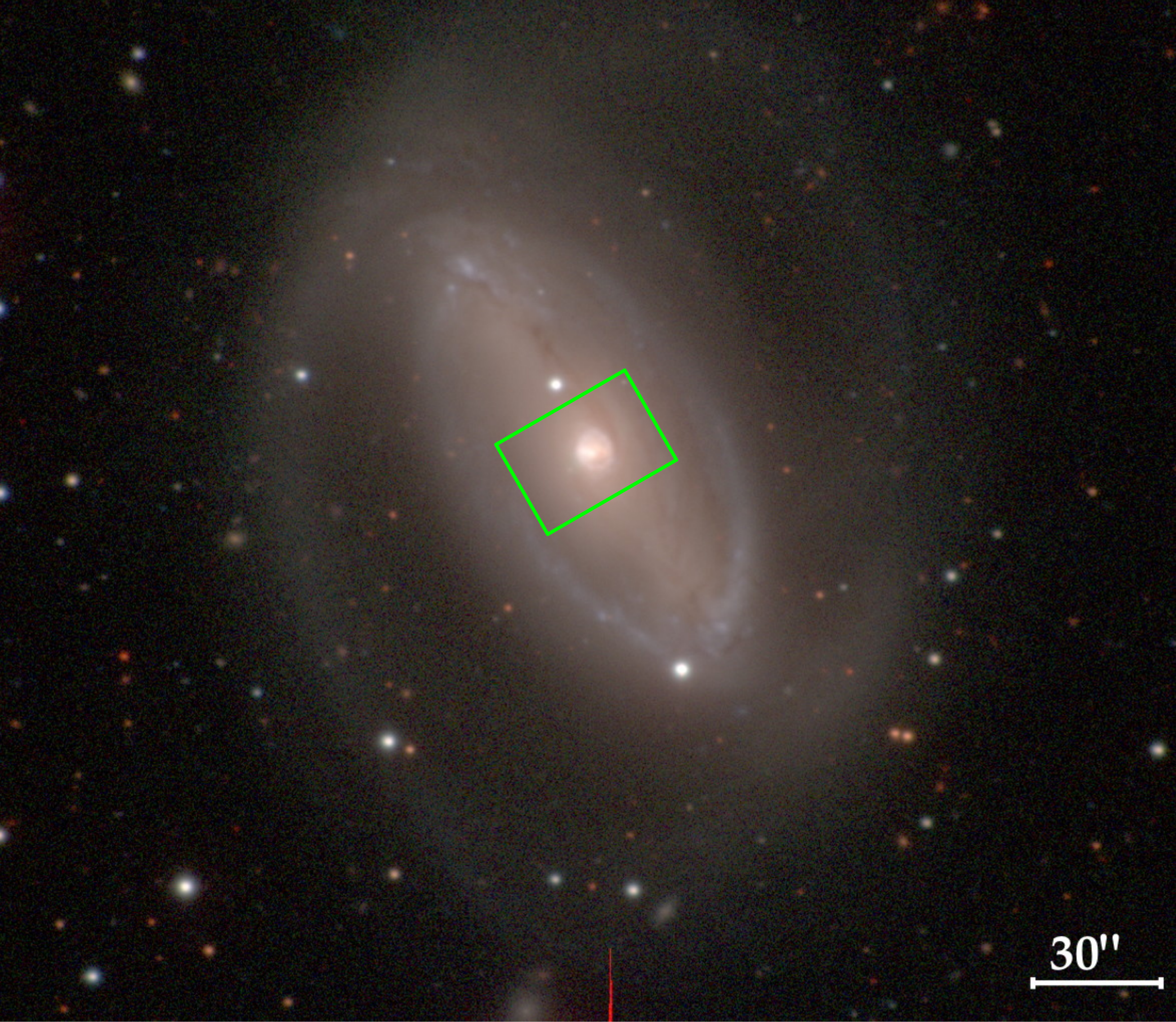}}
\subfloat[]{\includegraphics[scale = 0.3476, clip = true, trim = 5 5 5 5]{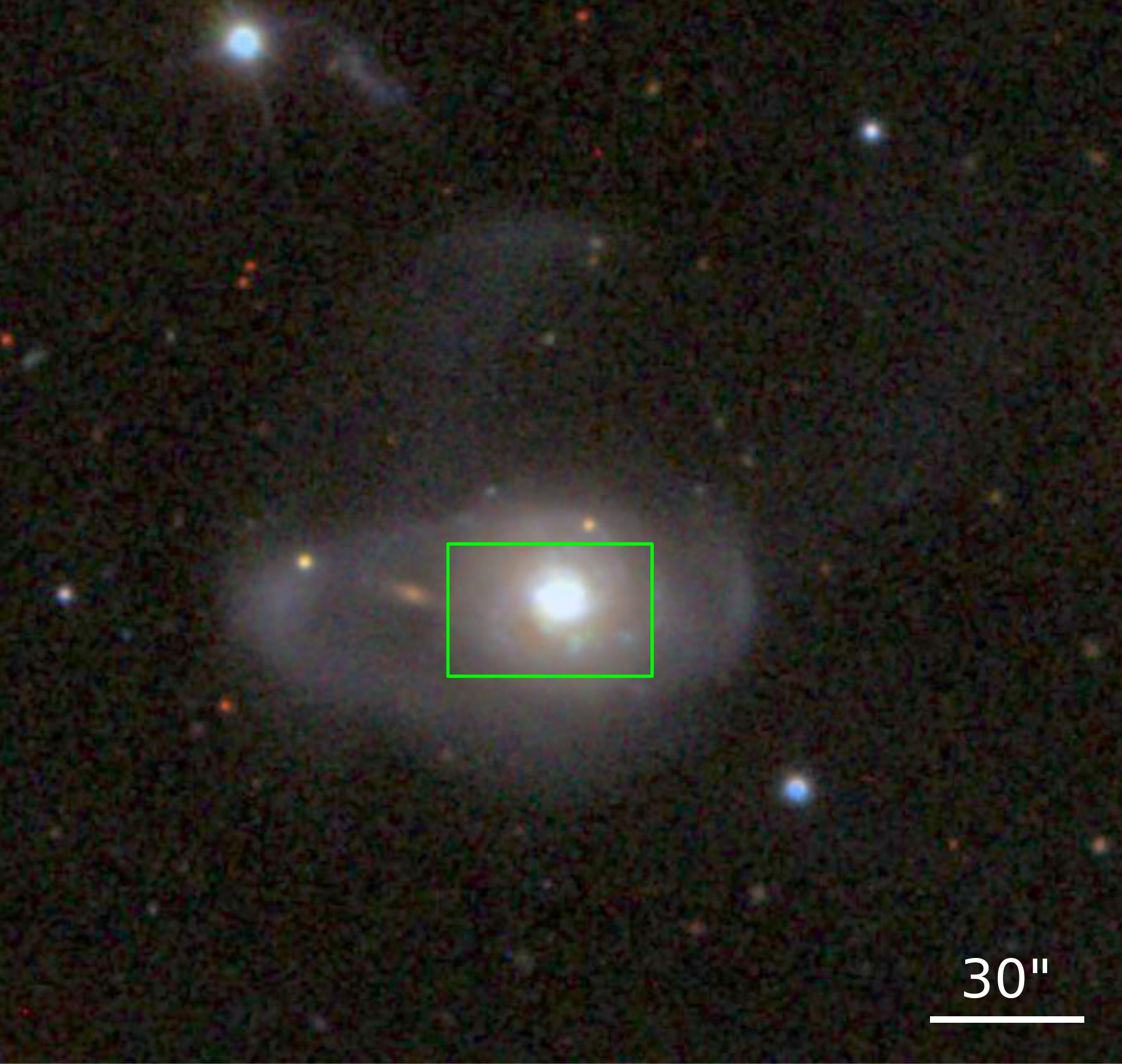}}
\caption{(Left) Carnegie-Irvine Galaxy Survey colour composite image of NGC~5728, and (right) SDSS colour composite image of NGC~7679. The \mbox{38$\times$25 arcsec$^2$} WiFeS footprint is overlaid in green on both images.}
\label{fig:dss}
\end{figure*}

\subsection{Component Selection}
\label{subsec:ANN}
We use an ANN called \textsc{The Machine} (Hampton et al., submitted) to determine which of the emission line fits (with 1, 2 or 3 Gaussian components) best represents the emission line spectrum of each spaxel. The ANN uses functions of a set of 86 input parameters to determine the most appropriate component classification. For each of the 1, 2 and 3 component fits, we input the signal-to-noise ratio (S/N) of the velocity dispersion and the S/N of the five strongest emission lines (\Ha~$\lambda$6563, \Hb~$\lambda$4861, \NII~$\lambda$6583, \mbox{\SII$\lambda \lambda$6717, 6731} and \OIII~$\lambda$5007) for each kinematic component, the S/N of the total flux (summed over all kinematic components) for the same five lines, and the reduced $\chi^2$ value of the emission line fit (output by \textsc{MPFIT}). The function coefficients for each of these input parameters were determined during the training stage. We provided the ANN with the values of the input parameters for a set of spaxels across 10 S7 galaxies, along with maps of the number of components required in each spaxel, determined by human astronomers through visual inspection. A cost function quantifies how `far' the calculated classifications are from the desired classifications and therefore allows the ANN to determine the most optimal set of coefficients.

Once training was complete, the coefficients were fixed and the ANN was run on the training galaxies to evaluate its success in selecting the correct number of components. The success of the neural network is quantified using the recall and the precision. The recall measures the percentage of spaxels to which the ANN assigns the correct number of components - for example, the percentage of spaxels assigned 1 component by the trainers which are also assigned 1 component by the ANN. The precision measures the percentage of spaxels to which the ANN correctly assigns a particular number of components - for example, the percentage of spaxels assigned 1 component by the ANN which were assigned 1 component by the trainers. The ANN performed with recalls of 94.1, 89.4 and 80.1 per cent for the 1, 2 and 3 component fits, respectively and precisions of 95.3, 89.4 and 71.0 per cent for the 1, 2 and 3 component fits, respectively. The ANN performs with high recalls and precisions, indicating that it is able to successfully determine the number of components required to represent the emission line spectra of the S7 galaxies. 

We also compare the performance of the ANN to the level of agreement between individual astronomers. We take the component maps produced by two of the trainers and construct a combined map containing information on only the spaxels with consistent component allocations between both trainers. We then calculate the recall and precision between this combined map and the component map produced by the third trainer. This comparison is performed for each trainer in comparison with the other two. We find recalls of 81.0-92.5, 39.9-95.9 and 62.4-96.2 per cent for the 1, 2 and 3 component fits, respectively and precisions of 67.5-98.1, 81.5-84.6 and 81.0-91.7 per cent for the 1, 2 and 3 component fits, respectively. The recalls and precisions of the ANN are consistent with or exceed the recalls and precisions between astronomers. The ANN therefore performs with an accuracy similar to visual inspection whilst also significantly reducing the time required to obtain component maps and providing reproducible results.

The high S/N (median \Ha\ and \OIII\ S/N of 50 and 20, respectively, for the spectra used in this study) and high spectral resolution of the S7 observations allow us to separate the observed line profiles into several kinematic components. However, high spectral resolution and/or high S/N observations become increasingly time demanding moving to fainter objects and/or higher redshifts. Constraining the relative intensities of multiple components in lower S/N and/or lower spectral resolution observations is non-trivial. In order to make our method directly applicable to the widest range of datasets, we base our decomposition on the total flux in each emission line in each spaxel rather than on the fluxes of individual kinematic components within those spaxels. We take the best emission line fit for each spaxel (determined by the ANN as described above) and sum the fluxes associated with each of the kinematic components to determine the total flux associated with each emission line. For the remainder of this paper we use these total fluxes rather than analysing each of the kinematic components individually. 

\subsection{Galaxy Sample}
This paper is based on S7 observations of two AGN host galaxies (NGC~5728 and NGC~7679) that both show clear mixing sequences between star formation and AGN activity on the \NIIHa\ vs. \OIIIHb\ diagnostic diagram (see Section \ref{subsec:decomp_data}). The presence of a clear mixing sequence is integral to our decomposition method, as discussed in Sections \ref{subsec:basis_spectra} and \ref{subsec:limitations}.

Figure \ref{fig:dss} shows a Carnegie-Irvine Galaxy Survey composite colour image of NGC~5728 and a SDSS composite colour image of NGC~7679, with the \mbox{38$\times$25 arcsec$^2$} WiFeS footprint overlaid in green. NGC~5728 was observed for 1800s with a typical seeing of 1.2 arcsec, corresponding to a physical resolution of $\sim$200~pc. NGC~7679 was observed for 3000s with a typical seeing of 1.8 arcsec, corresponding to a physical resolution of $\sim$300~pc. We achieve a continuum S/N of 30 and 40 per spaxel (at $\lambda \sim$~4900\AA) in the nuclei of NGC~5728 and NGC~7679, respectively. 

In the following sections we present and discuss flux maps and representative spectra of NGC~5728 and NGC~7679 to establish the presence of both current star formation and AGN activity in these two galaxies.

\subsubsection{NGC~5728}
NGC~5728 is an Sa galaxy which lies at a distance of 30.8 Mpc. It hosts a Seyfert 2 nucleus with a bolometric luminosity of \mbox{$\log$(L$_{bol}$/$L_\odot$) = 43.6} \citep{Vasudevan10}. Active star formation has been observed in the circumnuclear regions of the galaxy \citep{Rubin80, Schommer88}. 

\begin{figure*}
\captionsetup[subfigure]{labelformat=empty}
\subfloat[]{\includegraphics[scale = 1, clip = true, trim =  0 90 0 0]{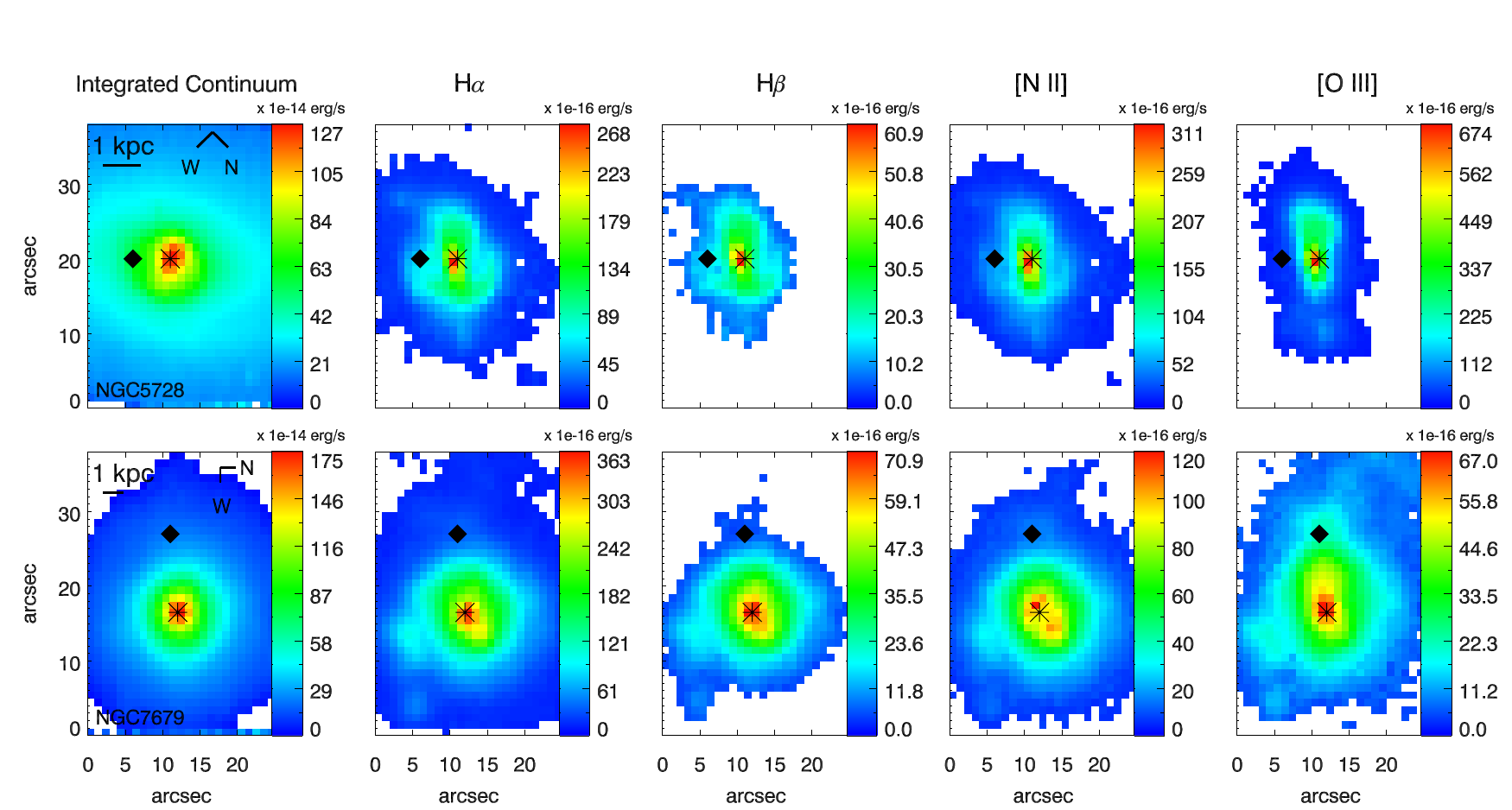}} \\[-0.5cm]
\subfloat[]{\includegraphics[scale = 0.9, clip = true, trim =  0 175 10 0]
{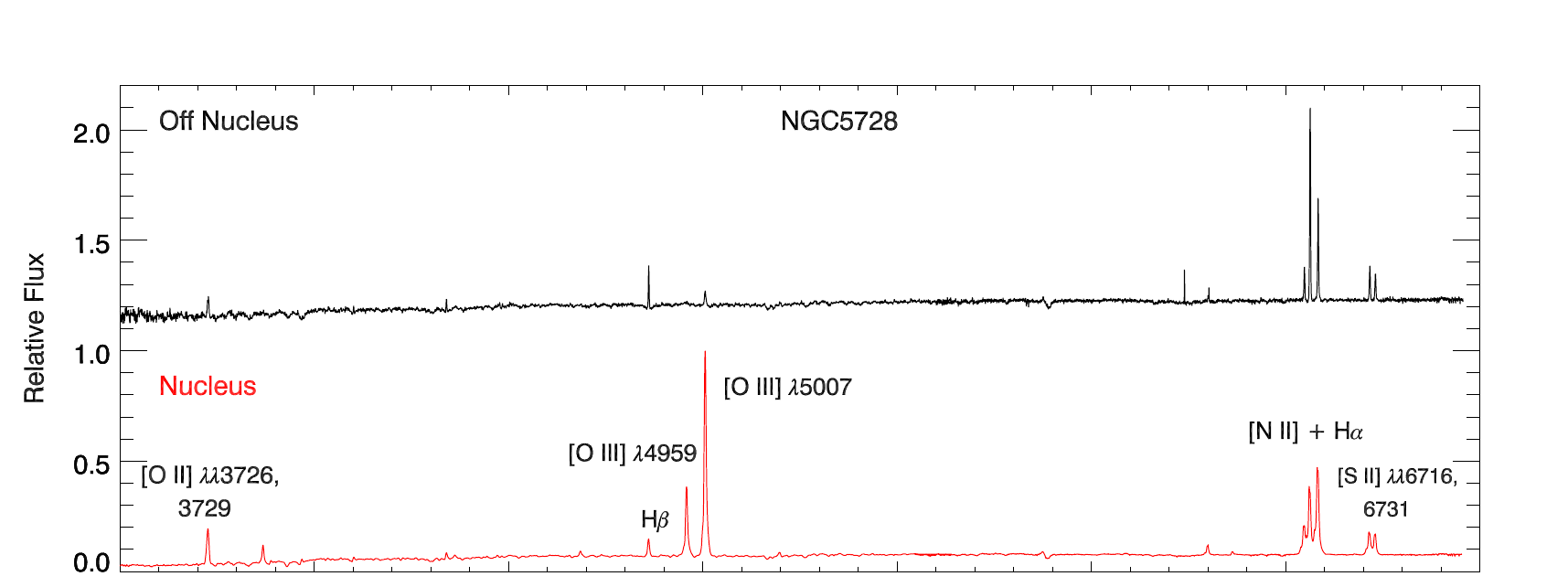}} \\[-1.4cm]
\subfloat[]{\includegraphics[scale = 0.9, clip = true, trim =  0 150 10 0]{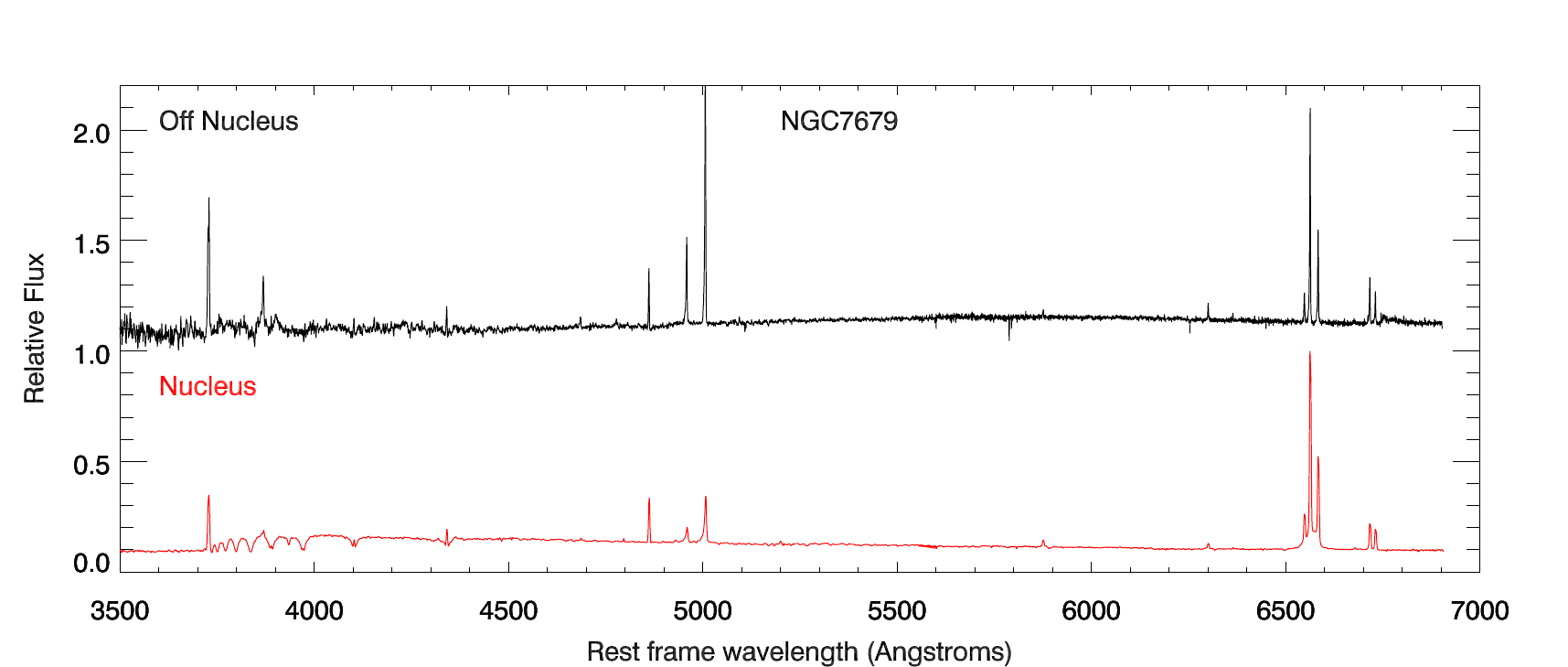}} \\[-0.3cm]
\caption{Top panels: Maps of the integrated continuum, \Ha, \Hb, \NII\ and \OIII\ emission across the WiFeS FOV for NGC~5728 and NGC~7679. We do not show integrated continuum measurements for spaxels with negative integrated continuum values. In each emission line map, we only show measurements for spaxels in which the relevant line is detected with \mbox{S/N $>$ 3}. Black asterisks on the images indicate the location of the galaxy centre (assumed to correspond to the peak of the integrated continuum emission). Black diamonds indicate the locations from which the off-nuclear spectra were extracted. Bottom panels: spectra extracted from 4 arcsecond diameter apertures centred on (red) nuclear and (black) off-nuclear regions of NGC~5728 and NGC~7679. }
\label{fig:galaxy_properties}
\end{figure*}

The top row of Figure \ref{fig:galaxy_properties} shows maps of the integrated continuum emission and the emission in each of the diagnostic lines (\Ha, \Hb, \NII\ and \OIII) across the WiFeS field of view (FOV) for NGC~5728. In the emission line maps, measurements are only shown for spaxels in which the relevant line is detected with \mbox{S/N $>$ 3}. The black asterisks mark the peak of the integrated continuum emission, assumed to be the galaxy centre. The continuum emission drops off fairly smoothly with radius, but the line emission (particularly the \OIII\ emission) is strongly asymmetric and appears to be enhanced along the north-west -- south-east axis (corresponding to a position angle of $\sim$120$^\circ$). \Hb\ is the weakest of the diagnostic emission lines but is still detected at \mbox{S/N $>$ 3} in 33 per cent of spaxels. When \Hb\ is detected, the average S/N is 13.0. In contrast, \Ha, \NII\ and \OIII\ are detected at an average S/N of 25.3, 30.9 and 22.8, respectively. Our S7 observations provide us with high S/N detections of the diagnostic emission lines across the nuclear region of NGC~5728.

The third row of Figure \ref{fig:galaxy_properties} shows spectra extracted from 4 arcsecond diameter apertures centred on (black) off-nuclear and (red) nuclear regions of NGC~5728. The off-nuclear spectrum is extracted from a region to the south-west of the galaxy nucleus (indicated by the black diamonds on the flux maps), offset from the axis of enhanced line emission. The off-nuclear spectrum is characterised by strong Balmer line emission (indicative of star formation), whereas the nuclear region is characterised by strong \NII\ and \OIII\ emission (indicative of AGN activity). These spectra confirm the presence of both active star formation and AGN activity within the WiFeS FOV for NGC~5728.

\subsubsection{NGC~7679}
\label{subsubsec:7679_properties}
NGC~7679 is a disturbed S0 galaxy and lies at a distance of 53.9 Mpc. The galaxy forms a wide pair with NGC~7682 (together Arp 216; \citealt{Arp66}) and is likely to have interacted with its companion in the past. The tidal features seen in Figure \ref{fig:dss} also suggest that the galaxy may have undergone a recent minor merger. NGC~7679 is classified as a Luminous Infrared Galaxy \citep{Armus09} and hosts a Seyfert 2 nucleus with a bolometric luminosity of \mbox{$\log$(L$_{bol}$/$L_\odot$) = 43.75} \citep{DellaCeca01}. The optical spectrum of NGC~7679 is strongly dominated by emission from \HII\ regions \citep[e.g.][]{Contini98}. 

The second row of Figure \ref{fig:galaxy_properties} shows maps of the integrated continuum emission and the emission in each of the diagnostic lines across the WiFeS FOV for NGC~7679. The spatial distribution of the continuum emission is very similar to the spatial distributions of the emission in all the lines except \OIII, which shows a clear enhancement on the eastern side of the galaxy. This enhancement in \OIII\ emission may be indicative of emission dominated by non-stellar sources. \Hb\ is detected in 61 per cent of the spaxels at an average S/N of 22.1. \Ha, \NII\ and \OIII\ are detected at average S/N of 49.8, 19.2 and 29.4, respectively. 

The bottom row of Figure \ref{fig:galaxy_properties} shows spectra extracted from 4 arcsecond diameter apertures centred on (black) off-nuclear and (red) nuclear regions of NGC~7679. The off-nuclear spectrum is extracted from the region to the east of the nucleus where the \OIII\ emission is enhanced relative to the continuum (indicated by the black diamonds on the flux maps). The off-nuclear spectrum shows strong \OIII\ emission (as expected), suggestive of AGN ionization. The \NII\ emission is relatively weak but this is likely to be indicative of a low gas phase metallicity in the AGN ionized gas. The nuclear spectrum shows strong Balmer line emission and clear stellar absorption features. We conclude that a significant proportion of the continuum emission is likely to be associated with A stars and that the majority of the line emission is associated with \HII\ regions. We emphasise that the AGN is not the dominant ionization mechanism in the nuclear region of NGC~7679.

\begin{figure*}
\centerline{\includegraphics[scale = 0.95, clip = true, trim =  0 200 0 0]{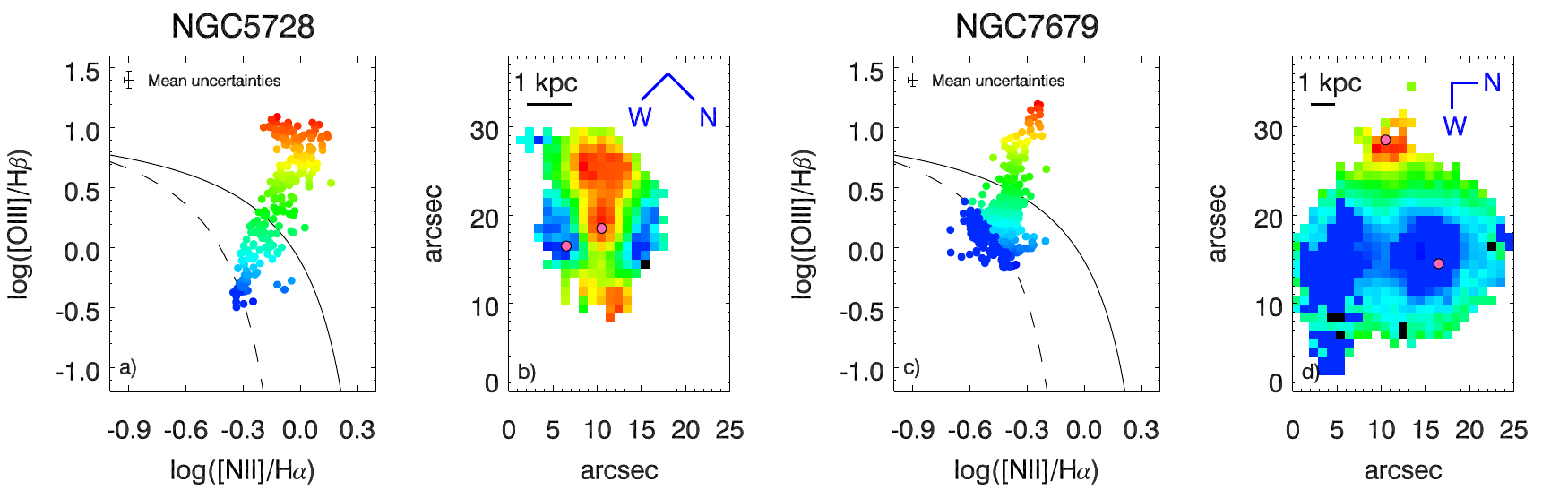}}
\caption{\NIIHa\ vs. \OIIIHb\ diagnostic diagrams populated with line ratios calculated using the total emission line fluxes of individual spaxels from the integral field datacubes of NGC~5728 (panel a) and NGC~7679 (panel c). The black asterisks indicate the line ratios of the galaxy nuclei, extracted from the nuclear spectra shown in Figure \ref{fig:galaxy_properties}. The solid black curves on the diagnostic diagrams delineate the \citet{Ke01a} theoretical upper bound to pure star formation, and all spectra above this line are dominated by AGN activity. The dashed black curves show the \citet{Ka03} empirical classification line, and all spectra below this line are dominated by star formation. Spectra lying between the two classification lines have significant contributions from both star formation and AGN activity. The spectra are colour-coded according to their projected distances along the line ratio sequences (in log-log space), and these colours are used to map the spatial variation in line ratios across the galaxies (panels b and d). All star formation dominated spaxels (lying below the Ka03 line) are coloured dark blue. The pink dots on the maps indicate the locations of the spaxels from which the basis spectra are extracted (see Section \ref{subsec:basis_spectra}).}
\label{fig:line_ratio_maps}
\end{figure*}

\subsubsection{Mixing Between Star Formation and AGN Activity}
\label{subsec:decomp_data}
NGC~5728 and NGC~7679 are selected to show clear evidence for mixing between star formation and AGN activity on the \NIIHa\ vs. \OIIIHb\ diagnostic diagram. Figure \ref{fig:line_ratio_maps} shows the diagnostic diagrams for NGC~5728 (panel a) and NGC~7679 (panel c), where each data point represents the line ratios calculated from the total \NII, \Ha, \OIII\ and \Hb\ fluxes of an individual spaxel. In this and all subsequent diagnostic diagrams we include only spectra in which all four diagnostic lines (\Ha, \Hb, \NII\ and \OIII) are detected to at least the 3$\sigma$ level. The black asterisks indicate the line ratios of the galaxy nuclei, extracted from the nuclear spectra shown in Figure \ref{fig:galaxy_properties}. The error bars in the top left corners of the diagnostic diagrams indicate the mean 1$\sigma$ uncertainties on the line ratios for each galaxy. The solid and dashed black curves on the diagnostic diagrams trace the \citet{Ke01a} and \citet{Ka03} classification lines (Ke01 and Ka03 lines), respectively. Spectra lying below the Ka03 line are dominated by emission associated with star formation, spectra lying above the Ke01 line are dominated by emission associated with more energetic ionization mechanisms (such as AGN activity or shocks), and spectra lying between the Ke01 and Ka03 lines have significant contributions from both emission associated with star formation and emission associated with more energetic ionization mechanisms.

Both NGC~5728 and NGC~7679 display smooth line ratio sequences spanning from the star-forming region to the AGN region of the diagnostic diagram. These sequences are likely to trace AGN fraction variations across the galaxies. We test this hypothesis by colour-coding the spectra according to their positions along the line ratio sequences, and using these colours to map the line ratio variations (panels b and d). For each galaxy the spectrum with the maximum \OIIIHb\ ratio is coloured red, and the spectrum with the minimum \OIIIHb\ ratio is coloured dark blue. The colours of all other spectra above the Ka03 line represent their projected distances along the line joining the two extreme spectra in $\log$(\NIIHa)-$\log$(\OIIIHb) space. All star formation dominated spectra (lying below the Ka03 line) are coloured dark blue.

We observe smooth variations in the line ratios across NGC~5728 and NGC~7679, providing strong evidence for spatially resolved mixing between star formation and AGN activity. The line ratios of NGC~5728 are elevated along the north-west -- south-east axis, and decrease smoothly to the north-east and south-west of the nucleus. The spectra in Figure \ref{fig:galaxy_properties} indicate that the emission in the nuclear region of NGC~5728 is dominated by AGN activity, whereas the region to the south-west of the nucleus is dominated by star formation. The decrease in line ratios towards the north-east and south-west of the nucleus is therefore likely to trace a decrease in the AGN fraction as emission from \HII\ regions becomes more prominent. The line ratios of NGC~7679 peak on the eastern side of the line-emitting region and decrease smoothly towards smaller galactocentric distances. The spectra in Figure \ref{fig:galaxy_properties} indicate that the eastern edge of the line-emitting region is dominated by AGN activity whereas emission from \HII\ regions is dominant in the centre of the galaxy. The decrease in line ratios moving towards the galaxy nucleus is therefore likely to trace a decrease in AGN fraction as \HII\ region emission becomes more significant. We find clear evidence for mixing between star formation and AGN activity in NGC~5728 and NGC~7679, making these ideal galaxies on which to demonstrate our new spectral decomposition method.

It is important to note that not all of the line ratio variations across NGC~5728 and NGC~7679 are attributable to mixing between star formation and AGN activity. We observe a separate line ratio sequence in the AGN region of the diagnostic diagram for NGC~5728, spanning from the high \OIIIHb\ end of the mixing sequence towards lower \NIIHa\ ratios. This line ratio sequence is likely to be driven by ionization parameter variations in the NLR gas \citep{Groves04}. We also observe a separate sequence within the star forming region of the diagnostic diagram for NGC~7679, spanning from the low \OIIIHb\ end of the mixing sequence towards lower \NIIHa\ ratios and higher \OIIIHb\ ratios (dark blue spaxels). This sequence approximately follows the SDSS \HII\ region sequence \citep[see e.g.][]{Ke06} and is therefore likely to trace metallicity variations between \HII\ regions \citep{Dopita86, Dopita00}. These metallicity variations may arise from chemical mixing associated with a previous minor merger event. The spectra along this line ratio sequence are all consistent with pure star formation and therefore it is not necessary to separate them into contributions from star formation and AGN activity.

\section{Separating Star Formation and AGN Activity}
\label{sec:method}

In Section \ref{subsec:decomp_data} we showed that there are clear relationships between the positions of spaxels on the diagnostic diagrams and their spatial locations within NGC~5728 and NGC~7679. These relationships are indicative of mixing between emission associated with star formation and emission associated with AGN heating across the NLRs (as discussed in \citealt{Davies14b, Davies14a}). Here, we build on this result and suggest that when clear mixing sequences are observed (such as in NGC~5728 and NGC~7679), the extinction corrected emission line fluxes (line luminosities) of the spectra along those mixing sequences can be reasonably approximated by linear superpositions of the line luminosities extracted from an \HII\ region basis spectrum and an AGN NLR basis spectrum. To test this, we first discuss how to obtain the basis spectra and how to separate emission associated with star formation and AGN activity in each spaxel of each galaxy (Sections \ref{subsec:basis_spectra} and \ref{subsec:method}), and then apply the decomposition method to our WiFeS data for NGC~5728 and NGC~7679 (Section \ref{sec:results}).

\begin{figure*}
\centerline{\includegraphics[scale = 1, clip = true, trim =  0 200 0 0]{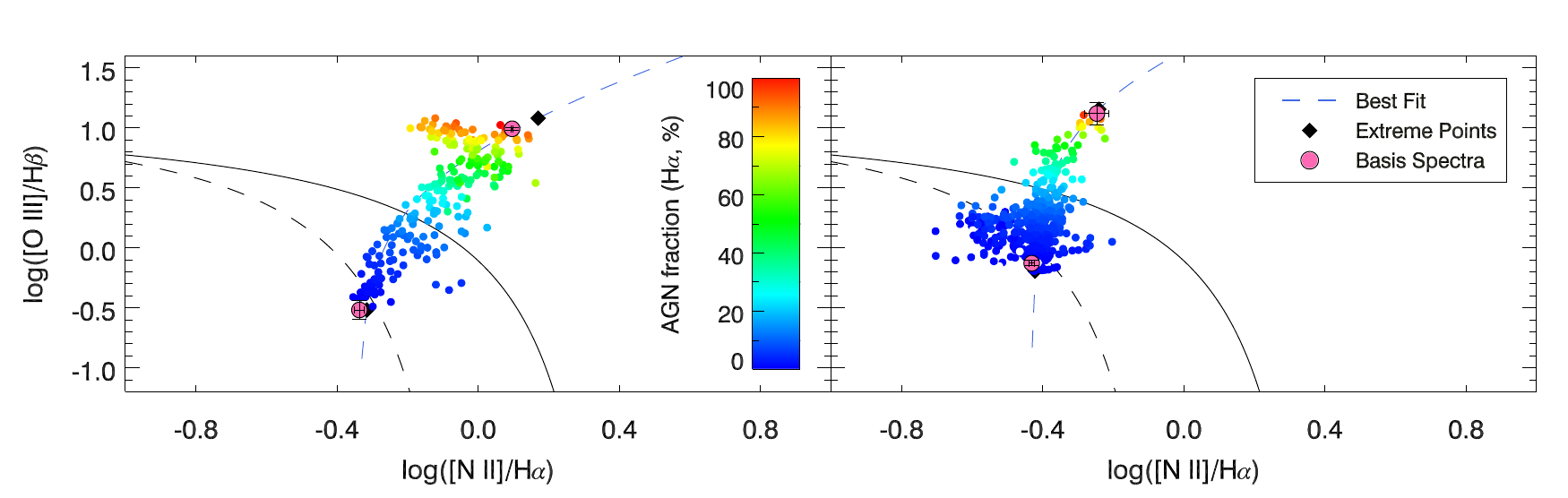}}
\caption{\NIIHa\ vs. \OIIIHb\ diagnostic diagrams for (left) NGC~5728 and (right) NGC~7679, with spectra colour-coded according to the calculated fraction of \Ha\ emission excited by AGN activity. The blue dashed curves show the best-fit starburst-AGN mixing curve for each galaxy. The `extreme points' of each mixing curve (black diamonds) are calculated by projecting the minimum and maximum observed \OIIIHb\ ratios onto the mixing curve. The optimal basis spectra for each galaxy (pink circles) are the observed spectra lying closest to each of the extreme points of the mixing curve. Error bars indicate the 1$\sigma$ errors on the line ratios of the basis spectra. The 1$\sigma$ error regions are sometimes smaller than the plotting symbols.}
\label{fig:basis_spectra}
\end{figure*}

\subsection{Selection of Basis Spectra}
\label{subsec:basis_spectra}
The clear mixing sequences seen in Figure \ref{fig:line_ratio_maps} suggest that many of the spectra in each galaxy fall approximately along a line connecting a single \HII\ region basis spectrum and a single AGN NLR basis spectrum (in linear-linear space). These basis spectra will be different for each mixing sequence, and can be obtained by fitting a line to the mixing sequence and selecting the spectra lying closest to the extreme points of this best fit line.

We use linear least squares minimisation in one dimension (with \NIIHa\ as the independent variable) to fit a straight mixing line to each \NIIHa\ vs. \OIIIHb\ mixing sequence in linear-linear space (blue dashed curves in Figure \ref{fig:basis_spectra}). The errors on the individual \NIIHa\ and \OIIIHb\ ratios are taken into account by the minimisation routine. The 1$\sigma$ scatters of the datapoints around the mixing lines for NGC~5728 and NGC~7679 are 0.09 dex and 0.06 dex, respectively. (These numbers are based on spectra above the Ka03 line only).

We define the maximal and minimal extreme points of each best-fit mixing curve (black diamonds) to be the points along the mixing curve with \OIIIHb\ ratios corresponding to the largest and smallest observed \OIIIHb\ ratios, respectively. The optimal \HII\ region and AGN NLR basis spectra for each mixing sequence (pink circles) are then the spectra with the line ratios closest to those of the minimal and maximal extreme points of the best-fit mixing curve, respectively. The \HII\ region and AGN NLR basis spectra for both galaxies lie in the pure star-forming and AGN dominated regions of the \NIIHa\ vs. \OIIIHb\ diagnostic diagram, respectively, and therefore the level of contamination due to mixing between ionization mechanisms in the basis spectra will be minimal. We note that the spatial resolution of the S7 observations (FWHM $\sim$~200-300~pc) is approximately an order of magnitude larger than the typical size of an \HII\ region, and therefore the \HII\ region basis spectrum is likely to be a luminosity-weighted superposition of many individual \HII\ region spectra. Similarly, the AGN NLR basis spectrum may be a superposition of emission from many AGN-ionized gas clouds within a single resolution element.   

The pink dots in Figure \ref{fig:line_ratio_maps} indicate the spatial locations of the spaxels from which the \HII\ region and AGN NLR basis spectra are extracted. The AGN NLR basis spectrum for NGC~5728 is extracted from the galaxy nucleus, and the \HII\ region spectrum is extracted from a region $\sim$~1~kpc west of the nucleus. The \HII\ region and AGN NLR basis spectra for NGC~7679 originate 1.6 and 4.3 kpc from the galaxy nucleus, respectively. The line ratio map for NGC~7679 indicates that the most AGN dominated regions are exclusively found at large galactocentric distances. Strong star formation swamps the AGN emission in the nuclear region of the galaxy, and the AGN only becomes dominant at the northern edge of the ionization cone where star formation is weak or absent (see Section \ref{subsubsec:7679_properties}). Consequently, the average S/N of the emission lines in the AGN NLR basis spectrum for NGC~7679 is significantly lower than the average S/N of the emission lines in the other three basis spectra (as illustrated by the error bars in Figure \ref{fig:basis_spectra}). The uncertainties on the basis spectra emission line fluxes are propagated through the decomposition process to produce the errors on the derived fluxes for the star formation and AGN components in each spaxel (see Section \ref{subsec:method}). 

We note that the uncertainties on the best fit mixing lines for NGC~5728 and NGC~7679 are negligible, and the selected basis spectra remain the same for every combination of the slope and intercept within the 1$\sigma$ ranges of these values. This would not be the case for galaxies with more complex line ratio distributions (such as those presented by \citealt{McElroy15}). We discuss caveats associated with applying our method to galaxies with complex line ratio distributions in Section \ref{subsec:limitations}.

\subsection{A New Spectral Decomposition Method}
\label{subsec:method}

The luminosity of any emission line $i$ in any spectrum $j$ which is a linear superposition of an \HII\ region basis spectrum and an AGN NLR basis spectrum can expressed as
\begin{equation}
L_{i}(j) = m(j) \times L_{i}(\textrm{SF}) + n(j) \times L_{i}(\textrm{AGN}) ~~(m(j), n(j) > 0)
\label{eqn:superposition}
\end{equation}
Here $L_{i}(\textrm{SF})$ and $L_{i}(\textrm{AGN})$ are the luminosities of the \HII\ region and AGN NLR basis spectra in emission line $i$. The coefficient $m(j)$ is the ratio of the SFR in spectrum $j$ to the SFR calculated from the \HII\ region basis spectrum. Similarly, $n(j)$ is the ratio of the luminosity of the AGN ionizing radiation field in spectrum $j$ to the luminosity calculated from the AGN NLR basis spectrum. If the superposition coefficients $m$ and $n$ can be calculated for all spectra along the mixing sequence, then it is trivial to determine the fractional contribution of the AGN to any emission line $i$ in any spectrum $j$ as follows:
\begin{equation}
f_{\textrm{AGN},i}(j) = \frac{n(j) \times L_{i}(\textrm{AGN})}{L_{i}(j)}
\label{eqn:fagn}
\end{equation}

\begin{figure*}
\centerline{\includegraphics[scale = 1, clip = true, trim =  0 110 0 0]{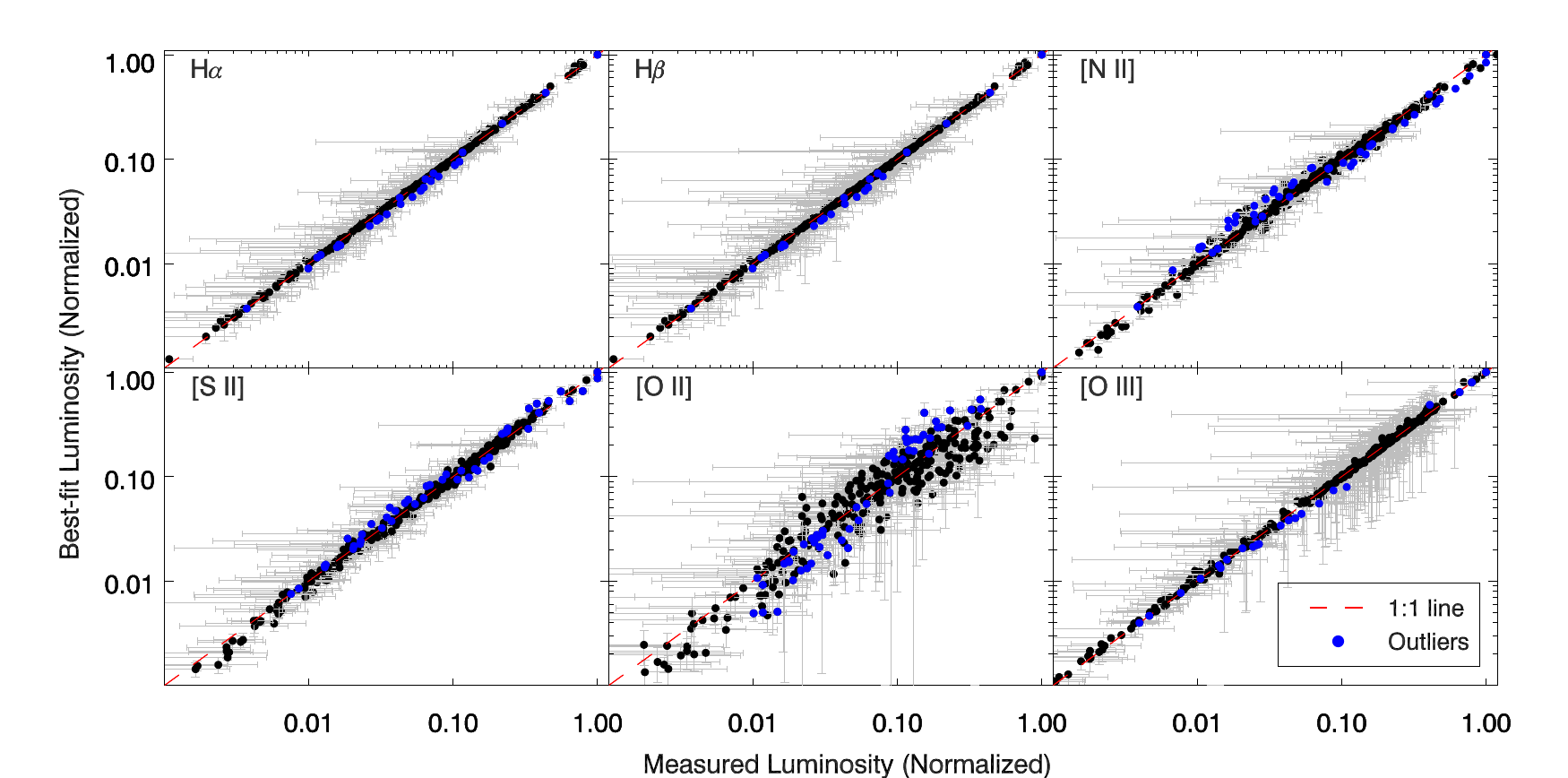}}
\caption{Comparison between the measured luminosities and the luminosities obtained from the best-fit linear superpositions of the basis spectra line luminosities for the six strongest lines in individual spaxels across NGC~5728 and NGC~7679. The luminosities have been normalised to the maximum measured luminosity in each emission line, and plotted on a log-log scale. The red dashed lines represent one-to-one correspondences between the measured and the best-fit luminosities. The blue points in each panel indicate the spectra which are not consistent with the one-to-one line within the error bars. The best-fit \Ha\ and \OIII\ luminosities are consistent with the measured luminosities (within the error bars) for 93 and 94 per cent of spaxels, respectively.}
\label{fig:residuals}
\end{figure*}

We constrain the superposition coefficients $m(j)$ and $n(j)$ for each spectrum by using the IDL routine \textsc{MPFIT} \citep{Markwardt09} to perform Levenberg-Marquardt least squares minimisation on Equation \ref{eqn:superposition} applied to the four strongest emission lines in our data (\Ha, \NII, \SII\ and \OIII). MPFIT simultaneously minimises over lines which are significantly separated in wavelength ($\sim$1725\AA\ between \OIII\ and \SII) and therefore it is essential to correct all fluxes (including the fluxes extracted from the basis spectra) for extinction before performing the least squares minimisation. We correct the measured emission line fluxes for extinction using the Balmer decrement, following the prescription outlined by \citet{Vogt13} and adopting the \citet{Fischera05} extinction curve with $R_V^A$~=~4.5. We assume an unreddened \Ha/\Hb\ ratio of 2.86 (appropriate for Case B recombination at \mbox{T = 10,000 K}; \citealt{Osterbrock06}). The median percentage errors on the Balmer decrements for NGC~5728 and NGC~7679 are 11 and 8 per cent, respectively. The errors on the Balmer decrements for individual spaxels are propagated through the extinction correction to produce the errors on the extinction corrected line fluxes. Once the extinction correction has been performed, the \Ha/\Hb\ ratio is constant and therefore including \Hb\ in the minimisation would not provide any extra constraint on the superposition coefficients. We normalise the basis spectra line luminosities to an \Ha\ luminosity of 1 to ensure that MPFIT primarily minimises over the relative rather than absolute line luminosities.

The minimisation algorithm takes into account both the errors on the line luminosities being fit and the errors on the basis spectra line luminosities. \textsc{MPFIT} returns uncertainties on each of the superposition coefficients, and these uncertainties are propagated to calculate uncertainties on the derived \HII\ region and AGN contributions. We emphasise that the superposition coefficients are calculated using only the relative luminosities of the \Ha, \NII, \SII\ and \OIII\ lines (rather than the entire spectra) and therefore the results of the decomposition are not sensitive to variations in the shape of the emission line profiles across the galaxies.

\section{Results}
\label{sec:results}
We apply the decomposition method presented in Section \ref{sec:method} to the mixing sequences seen in our WiFeS data for NGC~5728 and NGC~7679. In this section, we present our results and discuss various tests of our decomposition method. We show that linear superpositions of the basis spectra line luminosities are able to reproduce the emission line luminosities of $>$~85 per cent of spectra along the two mixing sequences. We also establish independent evidence of the success of the decomposition method by comparing the decomposed emission line maps to tracers of star formation and AGN activity at other wavelengths.

\subsection{Accuracy of Recovered Luminosities}
We use the determined superposition coefficients ($m(j)$ and $n(j)$) to calculate the \Ha, \Hb, \NII, \SII, \OIII\ and \OII\ luminosities associated with star formation and AGN activity in each spectrum along the mixing sequences of NGC~5728 and NGC~7679. We emphasise that once the superposition coefficients have been calculated, they can be used to decompose the emission in \textit{any} line which is robustly detected (with~S/N $>$~3) in both basis spectra.

\begin{figure*}
\captionsetup[subfigure]{labelformat=empty}
\subfloat[]{\includegraphics[scale = 0.7, clip = true, trim =  0 120 130 0]{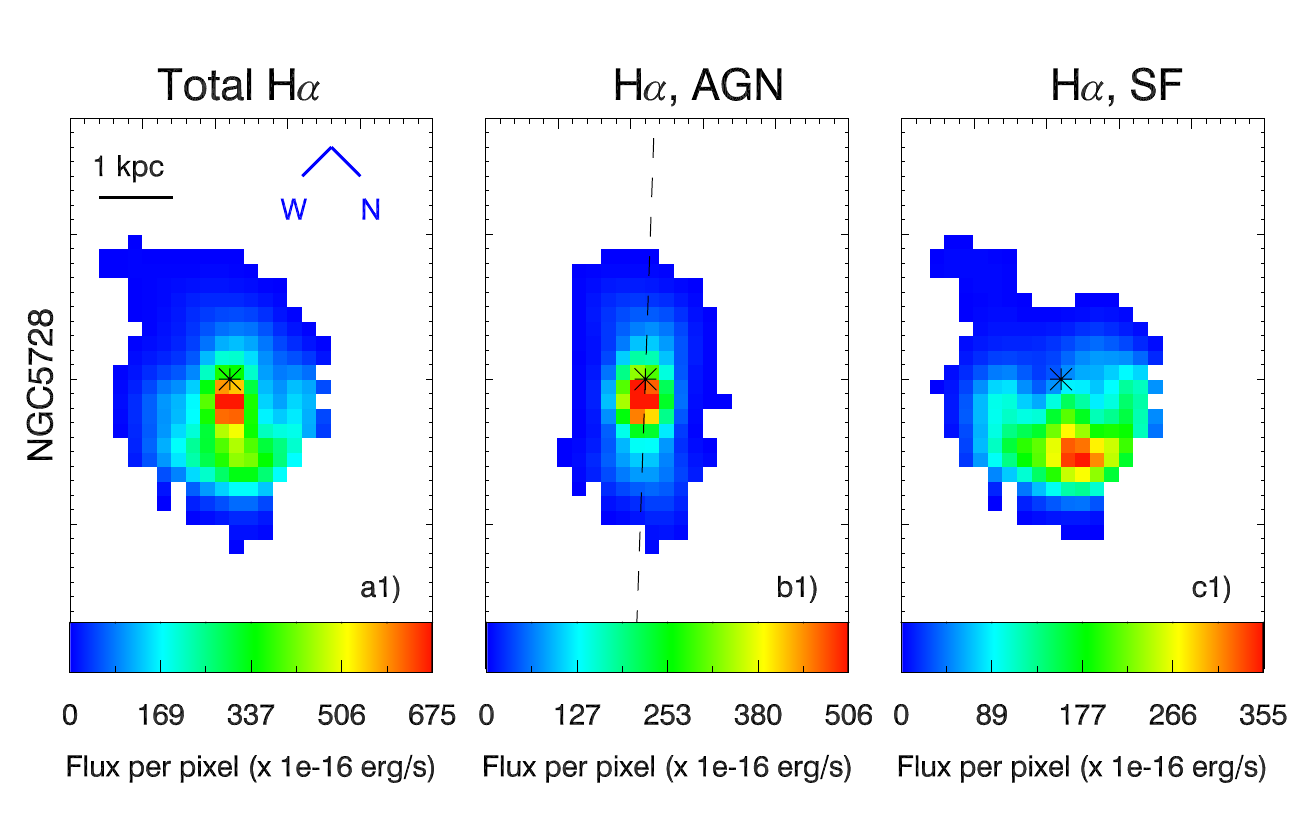}} 
\subfloat[]{\includegraphics[scale = 0.7, clip = true, trim =  0 120 130 0]{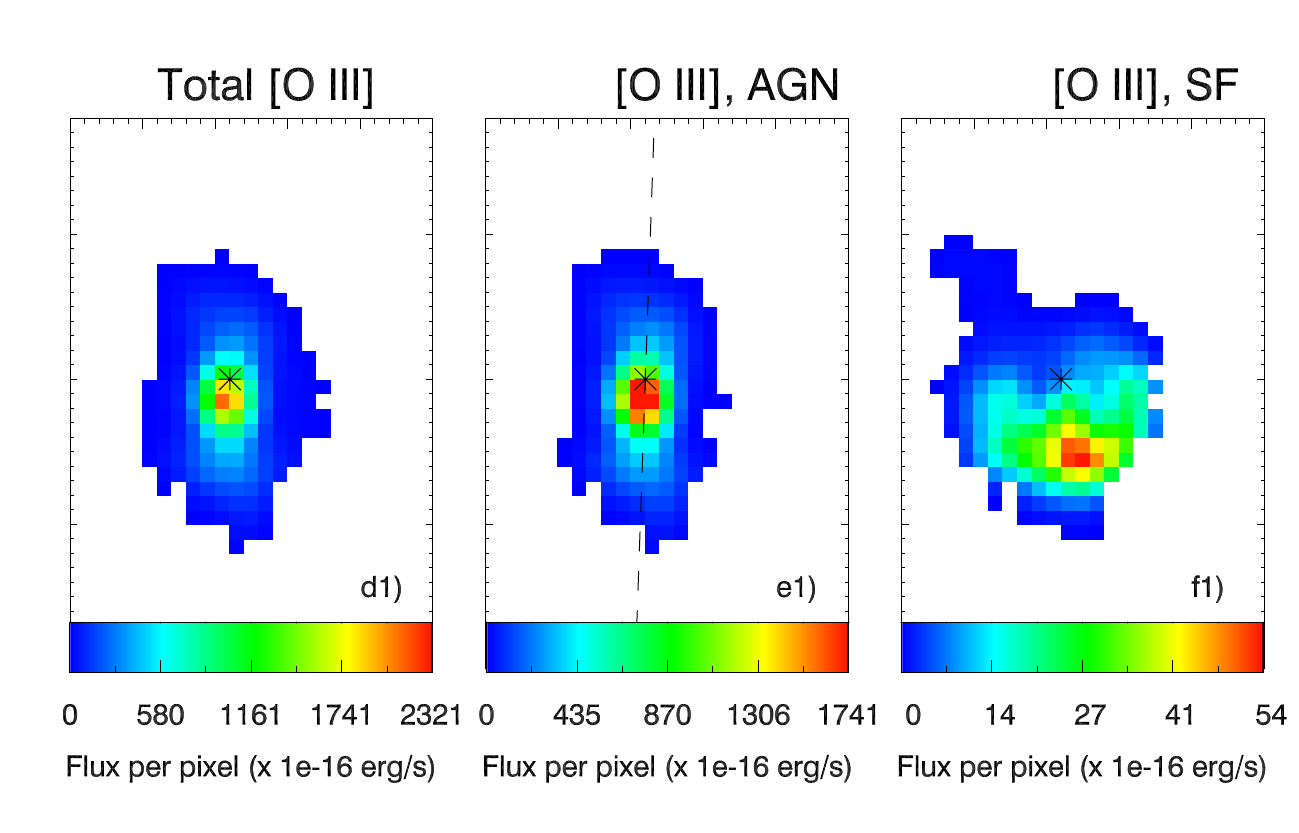}} \\[-1.8cm]
\subfloat[]{\includegraphics[scale = 0.7, clip = true, trim =  0 120 130 0]{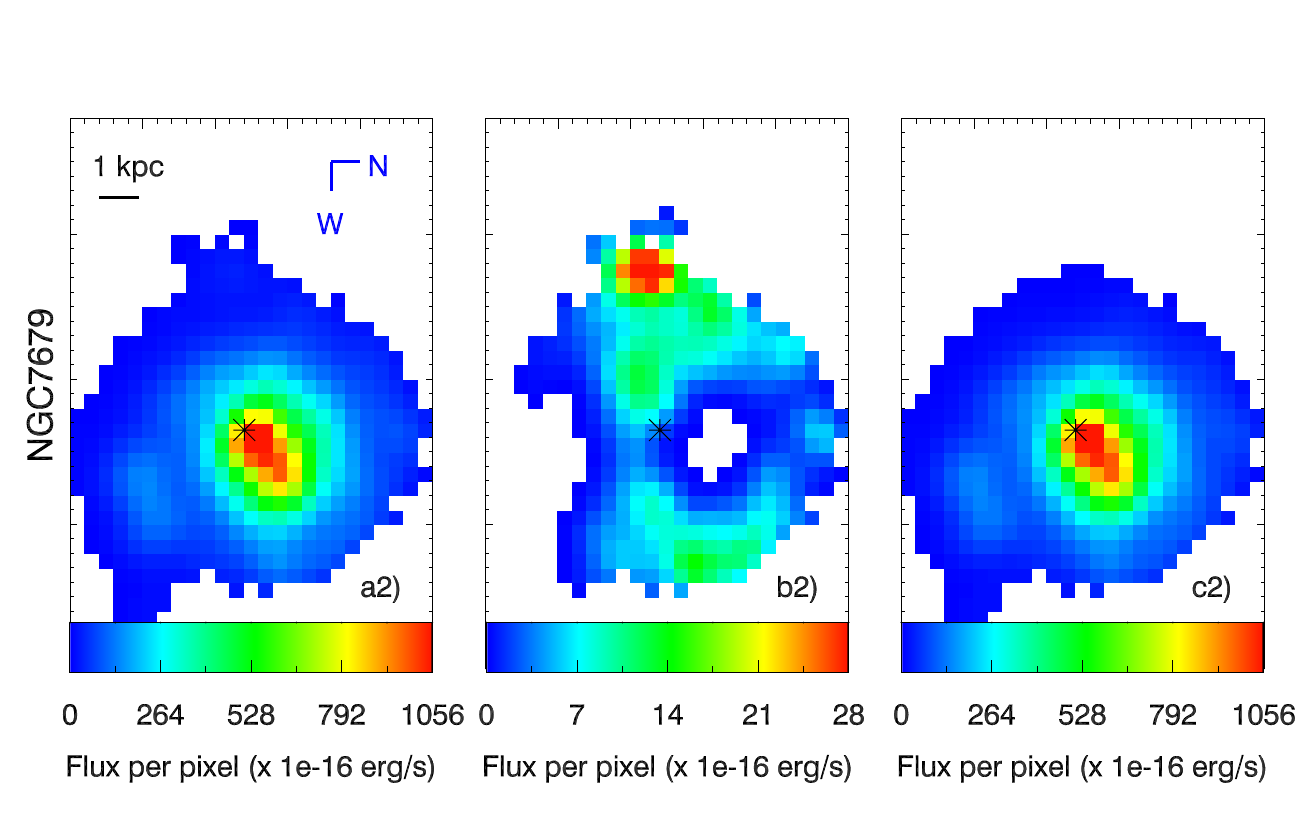}} 
\subfloat[]{\includegraphics[scale = 0.7, clip = true, trim =  0 120 130 0]{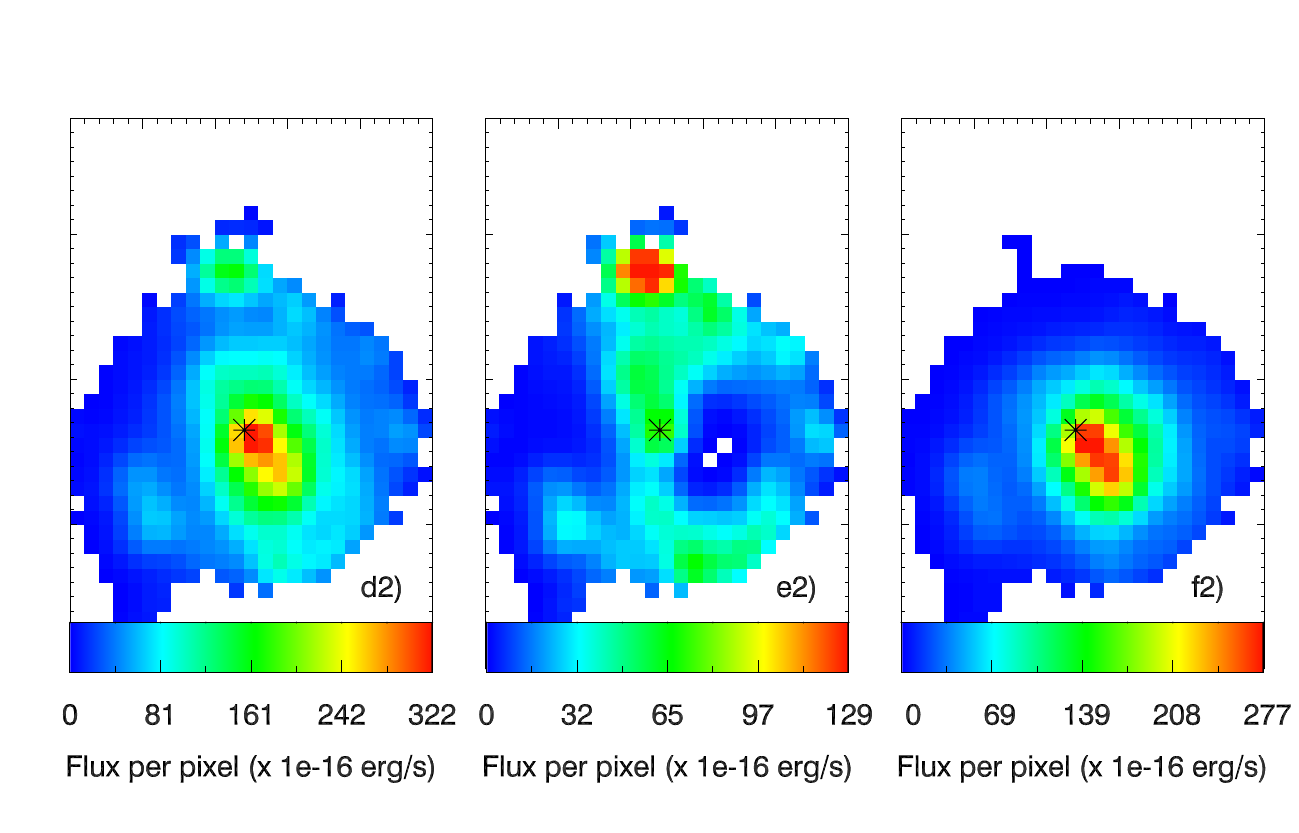}} \\[-0.8cm]
\caption{Maps of the total (extinction-corrected) \Ha\ and \OIII\ emission, and the \Ha\ and \OIII\ emission associated with star formation and AGN activity in (top) NGC~5728 and (bottom) NGC~7679. Black asterisks indicate the locations of the galaxy centres. The black dashed lines in panels b1) and e1) indicate the position angle of the AGN ionization cone identified by \citet{Wilson93} from HST imaging of NGC~5728. }
\label{fig:component_decomposition}
\end{figure*}

We find that the luminosities of individual emission lines in individual spectra along the mixing sequences of NGC~5728 and NGC~7679 are reasonably reproduced by linear superpositions of the line luminosities extracted from the optimal \HII\ region and AGN NLR basis spectra. Figure \ref{fig:residuals} compares the measured luminosities of the six strongest emission lines in the spectra of NGC~5728 and NGC~7679 to the `best-fit' luminosities (obtained from the best-fit linear superpositions of the basis spectra line luminosities; $L_{i}(j)$ in Equation \ref{eqn:superposition}). The error bars indicate the 1$\sigma$ errors which have been propagated through all steps of the analysis. The red dashed lines represent one-to-one correspondences between the measured and best-fit luminosities. The blue points in each panel indicate spectra that are not consistent with the one-to-one line within the error bars. The best-fit \Ha, \Hb, \NII, \SII\, \OII\ and \OIII\ luminosities are consistent with the measured luminosities (within the 1$\sigma$ errors) for 93, 94, 88, 88, 86 and 94 per cent of spectra, respectively. We note that we recover the observed \OII\ luminosities of the majority of spaxels, despite the fact that \OII\ is not used to constrain the superposition coefficients. The large scatter around the one-to-one line in the \OII\ panel is primarily driven by uncertainties in the extinction correction. (\mbox{\OII$\lambda \lambda$3727, 3729} is the shortest wavelength line used in our analysis and is therefore the most attenuated). We conclude that it is valid to approximate the luminosities of emission lines in spectra along the mixing sequences as linear superpositions of the line luminosities extracted from the optimal \HII\ region and AGN NLR basis spectra.  

\subsection{Emission Associated with Star Formation and AGN Activity}

\subsubsection{Relationship Between Line Ratios and AGN Fraction}
The most basic test of the success of the decomposition is to check that the spaxels with the largest and smallest line ratios have the largest and smallest calculated AGN fractions, respectively. The spectra in Figure \ref{fig:basis_spectra} are colour-coded according to their \Ha\ AGN fractions (calculated using Equation \ref{eqn:fagn}). The AGN fraction is largest for the spaxels with the largest \OIIIHb\ ratios and decreases smoothly towards smaller \NIIHa\ and \OIIIHb\ ratios, confirming that the least squares fitting algorithm used to determine the superposition coefficients behaves as expected. 

\subsubsection{NGC~5728}
\label{subsec:5728}
The top row of Figure \ref{fig:component_decomposition} shows the spatial distributions of the total \Ha\ and \OIII\ emission, and the \Ha\ and \OIII\ emission associated with star formation and AGN activity across NGC~5728. Black asterisks indicate the galaxy centre. The decomposed emission line images reveal the presence of two distinct line emitting structures. The \Ha\ and \OIII\ emission excited by the AGN ionizing radiation field delineates an ionization cone. The PA of the ionization cone (measured from the \Ha\ image) is $\sim$~120$^\circ$, consistent with the PA of 118$^\circ$ (indicated by the dashed black line in panels b1 and e1) measured from \emph{Hubble Space Telescope} (HST) imaging of NGC~5728 \citep{Wilson93}.

\begin{figure}
\centerline{\includegraphics[scale = 0.9, clip = true, trim = 40 175 240 35]{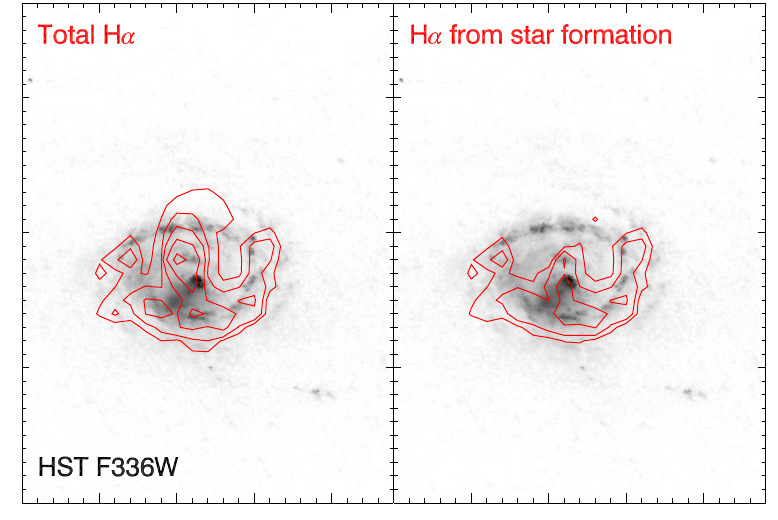}}
\caption{Zoom-in on the central 38$\times$25 arcsec$^2$ region of an HST WFC3 F336W image of NGC~5728, with contours of the (left) total \Ha\ emission and (right) \Ha\ emission due to star formation overlaid in red. The contour levels represent 5, 10, 20, 40 and 80 per cent of the maximum total \Ha\ flux per spaxel.}
\label{fig:NGC5728_overlay}
\end{figure}

The \Ha\ and \OIII\ emission excited by star formation traces the well known nuclear star forming ring which is thought to coincide with the Inner Lindblad Resonance of NGC~5728 \citep[see e.g.][]{Rubin80, Schommer88}. Figure \ref{fig:NGC5728_overlay} shows a HST Wide Field Camera 3 (WFC3) F336W image of the galaxy, with contours of the (left) total \Ha\ emission, and (right) \Ha\ emission associated with star formation, overlaid in red. The F336W filter covers \mbox{$\sim$310-360~nm} and is expected to be dominated by stellar continuum emission in regions with active star formation. The peak of the total \Ha\ emission is associated with the AGN ionization cone and does not coincide with a region of strong F336W emission. In contrast, the \Ha\ emission associated with star formation clearly traces the star forming ring seen in the F336W image. The star formation and AGN components recovered by the decomposition method have distinct spatial structures which are consistent with structures observed in HST images of NGC~5728, suggesting that our decomposition allows us to separate emission associated with different physical processes.

We can quantitatively test the success of our decomposition by using the decomposed emission line maps to calculate the SFR of the star forming ring and the bolometric luminosity of the AGN ($L_{bol, AGN}$), and then comparing these to independent estimates of the SFR and $L_{bol, AGN}$ from other wavelengths. We calculate the SFR of the ring using the \Ha\ emission attributed to star formation (\Ha$_{SF}$), and compare this to the SFR calculated from the Spitzer InfraRed Array Camera (IRAC) 8$\mu$m luminosity. The 8$\mu$m band covers strong polycyclic aromatic hydrocarbon (PAH) features at 7.7$\mu$m and 8.6$\mu$m which are excited by radiation from young massive stars and therefore trace active star formation. The 8$\mu$m emission can also be contaminated by stellar continuum emission, but this contamination can be removed using a 3.6$\mu$m band image \citep{Helou04}. We use convolution kernels from \citet{Aniano11} to convolve the 3.6$\mu$m and \Ha$_{SF}$ images to the same PSF as the 8$\mu$m image, and then calculate the total \Ha$_{SF}$ flux and the total 3.6$\mu$m and 8$\mu$m flux densities within the \mbox{25$\times$38 arcsec$^2$} WiFeS FOV. We correct the 8$\mu$m flux density for the stellar continuum contribution using the procedure presented by \citet{Helou04}, assuming \mbox{$\beta = 0.25$}. We convert the \Ha$_{SF}$ and corrected 8$\mu$m luminosities to SFRs using the calibrations of \citet{Kennicutt98} and \citet{Roussel01}, respectively. 

The SFRs implied by the \Ha$_{SF}$ and 8$\mu$m luminosities are 1.38$\pm$0.03 and \mbox{1.48$\pm$0.01 M$_\odot$ yr$^{-1}$}, respectively. The \Ha$_{SF}$ SFR is expected to be lower than the 8$\mu$m SFR because dust may obscure all of the optical light emitted from some regions of the galaxy. The two star formation rate estimates differ by only \mbox{0.1$\pm$0.03 M$_\odot$ yr$^{-1}$} and are therefore broadly consistent with one another. In contrast, the SFR calculated from the total \Ha\ luminosity (before separation into star forming and AGN components) is \mbox{2.00$\pm$0.04 M$_\odot$ yr$^{-1}$} which is \mbox{0.52$\pm$0.04 M$_\odot$ yr$^{-1}$} larger than the 8$\mu$m SFR. The AGN is responsible for a significant fraction of the \Ha\ emission across probed region of NGC~5728, and therefore the total \Ha\ luminosity is not reliable indicator of the SFR. Correcting the \Ha\ emission for the contribution of the AGN significantly improves the accuracy of the SFR calculated for the star forming ring of NGC~5728.

We calculate the bolometric luminosity of the AGN using the \OIII\ luminosity associated with AGN accretion ($L_{[O III], AGN}$), and compare this to the $L_{bol, AGN}$ calculated from the \textit{Swift} Burst Alert Telescope (BAT) 2-10 keV (hard X-ray) luminosity. The \OIII\ luminosity attributable to AGN accretion is \mbox{(4.31$\pm$0.08)$\times$10$^{41}$ erg s$^{-1}$}, which corresponds to a bolometric luminosity of \mbox{$\log(L_{bol, AGN}$/erg s$^{-1}$) = 43.8} (using a bolometric correction factor of 142; \citealt{Lamastra09}). This bolometric luminosity is broadly consistent with the value calculated from the \mbox{2-10 keV} luminosity (\mbox{$\log(L_{bol, AGN}$/erg s$^{-1}$) = 43.5-43.7}; \citealt{Vasudevan10}). The AGN is responsible for 94 per cent of the \OIII\ emission across the observed region of NGC~5728 and therefore calculating $L_{bol, AGN}$ using the total \OIII\ emission (before decomposition) yields very similar results.

Our decomposition method allows us to separate the contributions of star formation and AGN activity to the line emission across the central \mbox{$\sim$3~$\times$~3.75 kpc$^2$} region of NGC~5728. The emission associated with star formation is distributed along the star-forming ring seen in the HST F336W image of NGC~5728, and the SFR derived from \Ha$_{SF}$ is roughly consistent with the SFR derived from the 8$\mu$m luminosity over the same region. The emission associated with AGN activity delineates an ionization cone, the PA of the ionization cone is consistent with the PA calculated from HST narrow-band imaging of NGC~5728, and the AGN bolometric luminosity calculated from \OIII$_{AGN}$ is roughly consistent with the bolometric luminosity from the 2-10 keV luminosity over the same region. The comparison between our decomposed emission line maps and multi-wavelength observations of NGC~5728 provides compelling evidence that our decomposition has been successful.

\subsubsection{NGC~7679}
\label{subsec:7679}
The bottom row of Figure \ref{fig:component_decomposition} shows the spatial distributions of the total \Ha\ and \OIII\ emission and the \Ha\ and \OIII\ emission associated with each of the ionization mechanisms across NGC~7679. The star formation component traces a disk of \HII\ regions, whereas the emission associated with AGN activity is distributed in an ionization cone. The peak flux of the AGN component is coincident with the line ratio peak seen in Figure \ref{fig:line_ratio_maps}. We did not find any measurements of the ionization cone PA in the literature to compare with the orientation of the emission seen in panels b2) and e2) in Figure \ref{fig:component_decomposition}. The star forming and AGN components of the \Ha\ and \OIII\ emission have distinct morphologies which trace different physical processes occuring in NGC~7679.

The peak fluxes of the star formation and AGN components of the \OIII\ emission differ by a factor of $\sim$2.2, indicating that star formation and AGN activity are both important sources of \OIII\ emission in NGC~7679. We investigate how well our decomposition method separates emission associated with star formation and AGN activity in NGC~7679 by calculating the AGN bolometric luminosity using the total \OIII\ emission and then only the \OIII\ emission attributable to AGN activity, and comparing these estimates to the bolometric luminosity calculated from X-ray observations. The total and AGN contributions to the \OIII\ luminosity across the WiFeS FOV are \mbox{(1.23$\pm$0.04)$\times$10$^{42}$ erg s$^{-1}$} and \mbox{(4.21$\pm$0.31)$\times$10$^{41}$ erg s$^{-1}$}, respectively, corresponding to bolometric luminosities of \mbox{(1.89$\pm$0.06)$\times$10$^{44}$ erg s$^{-1}$} and \mbox{(5.98$\pm$0.44)$\times$10$^{43}$ erg s$^{-1}$}, respectively (using a bolometric correction factor of 142; \citealt{Lamastra09}). The bolometric luminosity calculated from the AGN component of the \OIII\ emission is consistent with the bolometric luminosity calculated from the X-ray luminosity (\mbox{5.6$\times$10$^{43}$ erg s$^{-1}$}; \citealt{DellaCeca01}) to within 1$\sigma$, whereas the total \OIII\ luminosity clearly over-estimates the AGN bolometric luminosity. Our decomposition allows us to isolate the contribution of the AGN to the \OIII\ emission of NGC~7679.

Star formation is the dominant source of \Ha\ emission in NGC~7679. The peak flux of the AGN component of the \Ha\ emission is a factor of $\sim$39 less than the peak flux of the star-formation component, and the spatial distributions of the total \Ha\ emission and the \Ha\ emission associated with star formation are very similar. Our decomposition indicates that star formation is responsible for 96 per cent of the \Ha\ emission in NGC~7679, consistent with the presence of many star formation dominated spaxels (see Figures \ref{fig:line_ratio_maps} and \ref{fig:basis_spectra}). The total \Ha\ luminosity associated with star formation within the WiFeS FOV is \mbox{2.68$\pm$0.02 erg s$^{-1}$}, corresponding to an SFR of \mbox{21.2$\pm$0.2 M$_\odot$ yr$^{-1}$}. In comparison, the SFR calculated from the 3.6$\mu$m and 8$\mu$m images (as described in \ref{subsec:5728}) is \mbox{11.35$\pm$0.6 M$_\odot$ yr$^{-1}$}, a factor of 1.9 (0.27 dex) lower than the SFR calculated from the \Ha\ emission. This discrepancy cannot be a result of under-correcting the \Ha\ emission for the contribution of the AGN, because 97 per cent of the \Ha\ emission originates from spaxels in the star forming and composite regions of the \NIIHa\ vs. \OIIIHb\ diagnostic diagram where the fractional contribution of the AGN to the \Ha\ emission is expected to be less than 30 per cent. The difference in the \Ha\ and 8$\mu$m star formation rates is likely to be driven by a combination of the difference in the timescales probed by the 8$\mu$m and \Ha\ emission \citep[see e.g.][]{Kennicutt12}, inaccuracies in accounting for dust extinction, and the 0.19 dex scatter in the \mbox{$L_{8\mu m}$-$L_{H\alpha}$} relation \citep{Roussel01}.

Decomposing the \Ha\ and \OIII\ emission of NGC~7679 into star forming and AGN components reveals an AGN ionization cone superimposed on a disk of \HII\ regions. The AGN bolometric luminosity calculated from the \OIII\ emission associated with AGN activity is consistent with the bolometric luminosity calculated from X-ray emission. The SFR calculated from the \Ha\ emission associated with star formation is a factor of 1.9 higher than the SFR calculated from the 8$\mu$m emission, but this discrepancy cannot be the result of under-correcting the \Ha\ emission for the contribution of AGN activity. Our decomposition method makes it possible to isolate the contributions of star formation and AGN activity to the optical emission of NGC~7679.

\section{Discussion}
\label{sec:discussion}
In this paper we have presented a simple method for determining the contributions of star formation and AGN activity to the luminosities of individual emission lines in integral field data of AGN host galaxies. The decomposition method can be applied to any galaxy with a clear starburst-AGN mixing sequence on the \NIIHa\ vs. \OIIIHb\ diagnostic diagram, and requires only the \Ha, \Hb, \NII, \SII\ and \OIII\ fluxes of the spectra along this mixing sequence. We have shown that both the morphologies of our decomposed emission line maps and the implied SFRs and AGN luminosities are consistent with independent tracers of star formation and AGN activity at other wavelengths, providing compelling evidence that our decomposition has been successful. 

The potential applications of our decomposition method are not limited to those presented in this paper. The superposition equation (Equation \ref{eqn:superposition}) can easily be extended to account for contributions from any number and type of ionization mechanisms, and in a future paper we will demonstrate that our method can be used to decompose spectra with significant contributions from star formation, shock excitation and AGN activity. Our decomposition technique can be applied to different combinations of emission lines, including some that are more easily accessible at higher redshifts. The \OIII, \OII\ and \Hb\ lines can be observed to at least z$\sim$0.5 with optical integral field spectrographs, and can be used to construct mixing sequences and determine the superposition coefficients for individual spaxels \citep{Davies14b}.

Our method builds on a body of techniques for identifying the underlying components contributing to the optical spectra of galaxies. In the following sections we outline some of these existing methods, discuss advantages and limitations of our method, and outline factors which could impact the inferred spatial distribution of the recovered star forming and AGN components.

\subsection{Comparison to Existing Methods}
Various techniques have been developed to quantify the relative contributions of different ionization mechanisms to single fibre optical emission line spectra of galaxies. The majority of these techniques are built on the assumption that the line luminosities of spectra in the composite region of the \NIIHa\ vs. \OIIIHb\ diagnostic diagram are linear superpositions of the line luminosities of \HII\ region and AGN NLR spectra. \citet{Heckman04} corrected the \OIII\ luminosities of observed composite spectra for the contribution of star formation by calculating the average contribution of the AGN to the \OIII\ luminosities of synthetic composite spectra in bins of \OIII\ luminosity. \citet{Ke06} suggested that the distances of galaxy spectra from the star-forming sequence of the diagnostic diagram (the`star-forming distance'; $d_{SF}$) could be used as a metric for the relative contribution of star formation to the line emission. \citet{Kauffmann09} used the positions of galaxy spectra along the AGN branch of the diagnostic diagram to estimate the fractional contribution of star formation to the \OIII\ luminosities of the galaxies. These pioneering techniques facilitated the first studies of the relationship between AGN properties and the properties of the host galaxy nuclei. The availability of integral field data has allowed us to extend existing decomposition methods and produce the first maps of the star formation rate and the strength of the AGN ionizing radiation field across AGN NLRs.

Techniques such as Mean Field Independent Component Analysis (MFICA) provide an alternate approach for analysing the underlying components of optical emission line spectra. \citet{Allen13} used MFICA to identify five components which can be linearly combined to accurately reproduce the emission line spectra of the majority of narrow emission-line galaxies in SDSS. Three of these components encompass the line ratio variations across the pure star forming galaxy population and the other two components account for contributions from harder ionization mechanisms. The emission line components do not contain sufficient information to quantify the relative contributions of star formation and AGN activity to the emission line spectra of individual galaxies, because the emission line spectra associated with pure star formation and pure AGN activity can have a range of different line ratios and therefore occupy multi-dimensional loci (rather than single locations) in the five dimensional MFICA space. However, MFICA offers a powerful avenue for understanding the sources of line ratio variation along the star forming sequence and AGN branch of the diagnostic diagram \citep[see e.g.][]{Richardson14}, and may subsequently be applied to separate star formation and AGN activity in individual galaxies.

\subsection{Limitations of Our Method}
\label{subsec:limitations}
\subsubsection{Shock Contamination and Variations in the ISM Conditions}
Our technique is based on the assumption that the observed line ratio variations are primarily driven by variations in the relative contributions of star formation and AGN activity. However, line ratio variations could also be driven by variations in the local ISM conditions and/or contamination from emission associated with other ionization mechanisms (such as shocks). Shock excited and AGN excited regions have similar diagnostic line ratios but can be differentiated using the ionized gas velocity dispersion, which is elevated in the presence of shocks \citep[e.g.][]{Rich11, Ho14}. This will be addressed in further detail in an upcoming paper. Significant variations in the ISM conditions would increase the scatter in the line ratios perpendicular to the mixing direction, ultimately leading to discrepancies between the measured line luminosities and the luminosities obtained from the best-fit linear superpositions of the basis spectra line luminosities. We emphasise that the results of the decomposition should only be used if the best-fit luminosities are consistent with the measured luminosities for the vast majority of spaxels. Ideally, all factors impacting the line ratios of AGN host galaxy spectra should be taken into account by performing the decomposition within a Bayesian framework, with priors informed by high resolution, multi-wavelength data. 

\subsubsection{Selection of Basis Spectra}
Our current method for selection of basis spectra cannot be applied to mixing sequences which do not span all the way from the star forming region to the AGN region of the \NIIHa\ vs. \OIIIHb\ diagnostic diagram (`truncated' mixing sequences), because one of the basis spectra would lie in the composite region of the diagnostic diagram and would therefore be significantly contaminated by mixing between ionization mechanisms. The basis spectra would not be sufficiently independent and the resulting decomposition would be unphysical. If none of the observed spectra lie in the star forming region of the diagnostic diagram, the \HII\ region basis spectrum could be chosen by intersecting the best-fit mixing curve with the SDSS \HII\ region sequence. We emphasise that the decomposition should only be performed with an \HII\ region basis spectrum lying in the star forming region of the diagnostic diagram and an AGN NLR basis spectrum lying in the AGN region of the diagnostic diagram.

\subsubsection{Application to Seyfert 1 Galaxies}
Our decomposition method only accounts for contributions from star formation and AGN NLR emission, and thus cannot be applied to Seyfert 1 galaxies in which emission from the broad line region (BLR) contributes significantly to the Balmer line emission. Further research is required to determine whether the addition of a third superposition coefficient (to account for the contribution of the BLR emission) would allow for accurate decomposition of Seyfert 1 galaxy spectra.

\subsection{Factors Impacting the Inferred Spatial Distributions of the Star Forming and AGN Components}
\label{subsec:dust_and_beam_smearing}
Comparing the spatial distributions of the emission associated with star formation and AGN activity may yield interesting insights into the role of AGN feedback in regulating star formation. However, in some situations the recovered spatial distributions of the star-forming and AGN components may not be representative of their intrinsic spatial distributions. Extreme dust attenuation in some regions of AGN host galaxies may impede the detection of the diagnostic emission lines and therefore prevent the underlying emission associated with star formation and/or AGN activity from being recovered. 

Beam smearing can also impact the inferred extent and spatial distribution of the star-forming and AGN components. The PSFs of our observations have FWHMs of 1.2 and 1.8 arcsec, corresponding to physical resolutions of 200 and 300 pc at the redshifts of our targets. These scales are approximately an order of magnitude larger than the typical sizes of \HII\ regions and AGN NLR clouds \citep[e.g.][]{Cecil02}, and therefore each observed spectrum is likely to be a luminosity-weighted superposition of the spectra of multiple \HII\ regions and/or AGN NLR clouds (as discussed in Section \ref{subsec:basis_spectra}). However, the star formation and AGN dominated regions in NGC~5728 and NGC~7679 are spatially separated by 1~kpc and 4~kpc, respectively, and therefore the AGN NLRs are well resolved in our observations. We conclude that the majority of the observed mixing is driven by the superposition of cospatial star formation and AGN dominated spectra, rather than by the smoothing of spectra from spatially distinct star forming and AGN dominated regions.

\section{Summary and Conclusions}
\label{sec:conc}
In this paper we have explained how emission associated with star formation and AGN activity can be separated both spatially and spectrally in integral field observations of AGN host galaxies. Our method is applicable to any galaxy with a clear mixing sequence on the \NIIHa\ vs. \OIIIHb\ diagnostic diagram, and requires only the \Ha, \Hb, \NII, \SII\ and \OIII\ fluxes for each spectrum along the mixing sequence. We demonstrated our method using integral field data for two AGN host galaxies from the S7 survey, NGC~5728 and NGC~7679. We showed that:
\begin{itemize}
\item Many of the spectra extracted from the datacubes of NGC~5728 and NGC~7679 lie along clear mixing sequences between star formation and AGN activity on the \NIIHa\ vs. \OIIIHb\ diagnostic diagram, and 
\item The emission line luminosities of $>$~85 per cent of the spectra along each mixing sequence can reproduced by linear superpositions of the emission line luminosities of one AGN dominated basis spectrum and one star formation dominated basis spectrum.
\end{itemize}

We separated the luminosity of each strong emission line in each spaxel into contributions from star formation and AGN activity, and compared our decomposed emission line maps to independent tracers of star formation and AGN activity at other wavelengths. 

The star formation component of NGC~5728 traces the star forming ring seen in the HST F336W image of this galaxy, and the SFR calculated from the star formation component of the \Ha\ emission is consistent with the SFR calculated from the 8$\mu$m emission over the WiFeS FOV. The AGN component is asymmetric and its alignment matches the position angle of the ionization cone identified in HST narrow band imaging of NGC~5728. The AGN bolomeric luminosity calculated from the AGN component of the \OIII\ emission also matches the bolometric luminosity calculated from the \mbox{2-10 keV} luminosity.

The star formation component of NGC~7679 traces \HII\ regions in a disk. The SFR calculated from the star formation component of the \Ha\ emission is a factor of 1.9 (0.27 dex) higher than the SFR calculated from the 8$\mu$m emission. This discrepancy cannot be due to under-correcting the \Ha\ emission for the contribution of the AGN, and may be driven by differences in the timescales traced by the different SFR indicators, inaccuracies in accounting for dust extinction, and/or intrinsic scatter in the SFR calibrations. The AGN component reveals a clear ionization cone, and the AGN bolometric luminosity calculated from the AGN component of the \OIII\ emission matches the bolometric luminosity calculated from X-ray observations of NGC~7679. Our decomposed emission line maps are consistent with independent tracers of star formation and AGN activity. We therefore conclude that our decomposition method allows us to robustly separate emission associated with star formation and AGN activity in NGC~5728 and NGC~7679.

The ability to separate emission associated with star formation and AGN activity will provide unique insights into the impact of AGN feedback on star formation in AGN host galaxies. The \Ha\ luminosities of the star formation component can be directly converted to SFRs, facilitating analysis of SFR gradients in AGN host galaxies. The spatial variations in the SFR surface densities can be compared to the spatial variations in the strength of the AGN ionizing radiation field to search for evidence of positive and/or negative AGN feedback (modulo the limitations of beam smearing and dust attenuation, as discussed in Section \ref{subsec:dust_and_beam_smearing}). Our results indicate that the combination of integral field spectroscopy and emission line ratio diagnostic diagrams is a powerful avenue by which to gain insights into the connection between star formation and AGN activity in galaxies.

\section{Acknowledgements}
We thank the anonymous referee for their detailed comments which improved the clarity of this manuscript. B.G. gratefully acknowledges the support of the Australian Research Council as the recipient of a Future Fellowship (FT140101202). M.D. and L.K. acknowledge the support of the Australian Research Council (ARC) through Discovery project DP130103925. J.S. acknowledges the European Research Council for the Advanced Grant Program Num 267399-Momentum. JKB acknowledges support from the Australian Research Council Centre of Excellence for All-sky Astrophysics (CAASTRO), through project number CE110001020.

\bibliography{decomposition_paper}

\end{document}